\documentclass[letterpaper,twocolumn,10pt]{article}
\PassOptionsToPackage{table}{xcolor}
\PassOptionsToPackage{hyphens}{url}
\usepackage{usenix2019_v3}

\usepackage{balance}
\usepackage{caption}
\usepackage{chngpage}
\usepackage{enumitem}
\usepackage{graphicx}
\usepackage{microtype}
\usepackage{subcaption}
\usepackage{xspace}

\microtypecontext{spacing=nonfrench}

\usepackage{textcomp}
\usepackage{upquote}

\emergencystretch=\maxdimen
\newenvironment{appxitem}{}{\vspace{\baselineskip}}

\usepackage[table]{xcolor}
\usepackage{adjustbox}
\usepackage{array}
\usepackage{multirow}

\newcommand*{\setuptable}{
	\renewcommand{\arraystretch}{1.1}
	\setlength{\arrayrulewidth}{0.1em}
	\setlength\tabcolsep{3.75pt}
	\centering
}

\newcommand*\rot{\rotatebox{90}}
\newcommand*{\headrow}[1]{\multicolumn{1}{c}{\adjustbox{angle=45,lap=\width-0.5em}{#1}}}
\newcommand*{\headline}[1][3cm]{\adjustbox{angle=45,lap=\width-.3mm}{\rule{#1}{0.075em}}}

\newcommand{\hc}{\cellcolor[gray]{1}}
\newcommand*{\headcol}[2]{\multirow{#1}{*}{\rot{#2}}}

\usepackage{wasysym}
\newcommand*{\full}{\CIRCLE}
\newcommand*{\prt}{\LEFTcircle}
\newcommand*{\none}{\Circle}
\newcommand*{\na}{}

\newcommand{\sig}{\cellcolor{green!10}}

\usepackage{pifont}
\newcommand*{\good}{{\textcolor[RGB]{112,173,71}{\ding{51}}}}
\newcommand*{\bad}{{\textcolor[RGB]{192,0,0}{\ding{55}}}}

\usepackage[disable]{todonotes}

\newcommand*{\anyone}[1]{\todo[inline,color=blue!25]{Anyone: #1}}

\usepackage{listings}
\usepackage{color}
\usepackage{lstautogobble} 

\definecolor{backgroundColor}{HTML}{fafafa}
\definecolor{keywordColor}{HTML}{d73a49}
\definecolor{stringColor}{HTML}{032f62}
\definecolor{commentColor}{HTML}{606770}
\definecolor{numberColor}{HTML}{D33682}
\colorlet{punctuationColor}{red!60!black}
\definecolor{scopingColor}{RGB}{20,105,176}

\lstdefinelanguage{csp}{
	upquote=true,
	otherkeywords={
		base-uri, block-all-mixed-content, child-src, connect-src, default-src, disown-opener, font-src, form-action, frame-ancestors, frame-src, img-src, manifest-src, media-src, navigate-to, object-src, plugin-types, referrer, report-sample, report-to, report-uri, require-sri-for, sandbox, script-src, strict-dynamic, style-src, upgrade-insecure-requests, worker-src
	},
	morestring=[b]",
	morestring=[b]',
	basicstyle={\small\ttfamily\color{commentColor}},
	stringstyle={\color{stringColor}},
	literate=*
	{;}{{{\color{punctuationColor}{;}}}}{1}
}

\colorlet{attributeColor}{yellow}
\lstdefinelanguage{HTML5}{
	language=html,
	sensitive=true, 
	alsoletter={<>=-},
	otherkeywords={
		<html>, <head>, <title>, </title>, <meta, />, </head>, <body>,
		<canvas, \/canvas>, <script>, </script>, </body>, </html>, <!, html>, <style>, </style>, ><
	},  
	ndkeywords={
		=,
		charset=, id=, width=, height=,
		border:, transform:, -moz-transform:, transition-duration:, transition-property:, transition-timing-function:
	},
	ndkeywordstyle={\color{attributeColor}},
	morecomment=[s]{<!--}{-->},
	tag=[s]
}

\lstdefinelanguage{JavaScript}{
	morekeywords={typeof, new, true, false, catch, function, return, null, catch, switch, var, if, in, while, do, else, case, break},
	morecomment=[s]{/*}{*/},
	morecomment=[l]//,
	morestring=[b]",
	morestring=[b]'
}

\lstdefinelanguage{json}{
	morestring=[b]",
	morestring=[b]',
}

\usepackage[all]{nowidow}

\begin{document}
	
\date{}

\title{That Was Then, This Is Now: A Security Evaluation of Password Generation, Storage, and Autofill in Browser-Based Password Managers\thanks{\textbf{This paper will appear at USENIX Security 2020.}}}

\author{
	{\rm Sean Oesch}\\
	University of Tennessee, Knoxville\\
	toesch1@vols.utk.edu
\and
	{\rm Scott Ruoti}\\
	University of Tennessee, Knoxville\\
	ruoti@utk.edu
}

\maketitle

\begin{abstract}
Password managers have the potential to help users more effectively manage their passwords and address many of the concerns surrounding password-based authentication.
However, prior research has identified significant vulnerabilities in existing password managers; especially in browser-based password managers, which are the focus of this paper.
Since that time, five years has passed, leaving it unclear whether password managers remain vulnerable or whether they have addressed known security concerns.
To answer this question, we evaluate thirteen popular password managers and consider all three stages of the password manager lifecycle---password generation, storage, and autofill.
Our evaluation is the first analysis of password generation in password managers, finding several non-random character distributions and identifying instances where generated passwords were vulnerable to online and offline guessing attacks.
For password storage and autofill, we replicate past evaluations, demonstrating that while password managers have improved in the half-decade since those prior evaluations, there are still significant issues; these problems include unencrypted metadata, insecure defaults, and vulnerabilities to clickjacking attacks.
Based on our results, we identify password managers to avoid, provide recommendations on how to improve existing password managers, and identify areas of future research.
\end{abstract}


\section{Introduction}
\label{sec:introduction}

Despite the well-established problems facing password-based authentication, it continues to be the dominant form of authentication used on the web~\cite{bonneau2012quest}.
Because passwords that are difficult for an attacker to guess are also hard for users to remember, users often create weaker passwords to avoid the cognitive burden of recalling them~\cite{dell2010password,riley2006password}.
In fact, with the increase in the number of passwords users are required to store, they often reuse passwords across websites~\cite{das2014tangled,florencio2007large,pearman2017let,wang2018end}.
Herley points out that this rejection of security advice by users is rational when the low percentage of users affected by breaches is contrasted with the effort required~\cite{herley2009so}.
However, the number of data breaches is on the rise~\cite{securityscorecard}, and this situation leaves many users vulnerable to exploitation.

Password managers can help users more effectively manage their passwords.
They reduce the cognitive burden placed upon the user by generating strong passwords, storing those passwords, and then filling in the appropriate password when a site is visited.
The user is now able to follow the latest security advice regarding passwords without placing a high cognitive burden on themselves.
But password managers are not impervious to attack.
Li et al.~\cite{li2014emperor} previously found significant vulnerabilities in major password managers like LastPass and RoboForm.
Both Silver et al.~\cite{silver2014password} and Stock and Johns~\cite{stock2014protecting} demonstrated that browser-based password managers, including LastPass and 1Password, are vulnerable to cross-site scripting attacks (XSS) and network injection attacks as a result of their password autofill features.

Since these studies five or more years have passed, leaving it unclear whether password managers remain vulnerable or whether they are now ready for broad adoption.
To answer this question, we update and expand on these previous results and present a thorough, up-to-date security evaluation of thirteen popular password managers.
We provide a comprehensive evaluation of browser-based password managers, including five browser extensions and six password managers integrated directly into the browser.
We also include two desktop clients for comparison.  

In our evaluation, we consider the full password manager lifecycle~\cite{choong2014cognitive}---password generation (Section~\ref{sec:generation}), storage (Section~\ref{sec:storage}), and autofill (Section~\ref{sec:autofill}).
For password generation, we evaluate a corpus of 147 million passwords generated by the studied password managers to determine whether they exhibit any non-randomness that an attacker could leverage.
Our results find several issues with the generated passwords, the most severe being that a small percentage of shorter generated passwords are weak against online and offline attacks (shorter than 10 characters and 18 characters, respectively).
We also replicate earlier work examining the security of password storage~\cite{gasti2012security} and autofill~\cite{li2014emperor,silver2014password,stock2014protecting}.

Our results find that while password managers have improved in the past five years, there are still significant security concerns.
We conclude the paper with several recommendations on how to improve existing password managers as well as identifying future work that could significantly increase the security and usability of password managers generally (Section~\ref{sec:discussion}).

Our \textbf{contributions} include: 

\begin{enumerate}[nosep]
	
	\item Our research finds that app-based and extension-based password managers have improved security compared to five years ago. However, there are still residual vulnerabilities that need to be addressed---for example, several tools will automatically fill passwords into compromised domains without user interaction and others that do require user interaction allow users to disable it.
	As such, it is important to both carefully select a password manager and to configure it properly, something that may be difficult for many users.
	
	\item To our knowledge, this paper is the first evaluation of password generation in password managers. As part of this evaluation, we generated 147 million passwords representing a range of different password managers, character composition policies, and length. We evaluated this corpus using various methods (Shannon entropy, $\chi^2$ test, zxcvbn, and a recurrent neural net) to find abnormalities and patterns in the generated passwords. We found several minor issues with generated passwords, as well as a more serious problem where some generated passwords are vulnerable to online and offline attacks.

	\item Our work is the most comprehensive evaluation of password manager security to date. It studies the largest number of password managers (tied with Gasti and Rasmussen\cite{gasti2012security}) and is the only study that simultaneously considers all three stages of the password manager lifecycle~\cite{choong2014cognitive}---password generation, storage, and autofill (prior studies considered either storage or autofill, but not both simultaneously).
	
	\item Prior security evaluations of password managers in the literature are now five or more years old. In this time, there have been significant improvements to password managers. In our work, we partially or fully replicate these past studies~\cite{gasti2012security,li2014emperor,silver2014password,stock2014protecting} and demonstrate that while many of the issues identified in these studies have been addressed, there are still problems such as unencrypted metadata, unsafe defaults, and vulnerabilities to clickjacking attacks.
\end{enumerate}

\section {Background}\label{sec:background}  
In this section, we describe the responsibilities of a password manager.
We also describe prior work that has analyzed password managers.

\subsection{Password Managers}
In the most basic sense, a password manager is a tool that stores a user's credentials (i.e., username and password) to alleviate the cognitive burden associated with a user remembering many unique login credentials~\cite{li2014emperor}.
This store of passwords is commonly referred to as a \emph{password vault}.
The vault itself is ideally stored in encrypted form, with the encryption key most commonly derived from a user-chosen password known as the \emph{master password}.
Optionally, the password vault can be stored online, allowing it to be synchronized across multiple devices.

In addition to storing user-selected passwords, most modern password managers can help users generate passwords.
Password generation takes as input the length of the desired password, the desired character set, and any special attribute the password should exhibit (e.g., at least one digit and one symbol, no hard to recognize characters).
The password generator outputs a randomly generated password that meets the input criterion.

Many password managers also help users authenticate to websites by automatically selecting and filling in (i.e., \emph{autofill}) the appropriate username and password.
If users have multiple accounts on the website, the password manager will allow users to select which account they wish to use for autofill.


If properly implemented and used, a password manager has several tangible benefits to the user:
\begin{enumerate}
	\item It reduces the cognitive burden of remembering usernames and passwords.
	\item It is easy to assign a different password to every website, addressing the problem of password reuse.
	\item It is easy to generate passwords that are resilient to online \emph{and} offline guessing attacks.
\end{enumerate}


\subsection{Related Work}
\label{subsec:relatedwork}
Several studies have looked at various aspects of password manager security.

\textbf{Web Security}
Li et al.~\cite{li2014emperor} analyzed the security of five extension-based password managers, finding significant vulnerabilities in the tools as well as the websites that hosted the user's password vault.
These vulnerabilities included logic and authorization errors, misunderstandings about the web security model, and CSRF/XSS attacks.
They also found that password managers that were deployed using bookmarklets did not use iframes properly, leaving the tools vulnerable to malicious websites.

Google's Project Zero found a bug in LastPass where credentials from the last visited site could be leaked to the currently visited site; this bug has since been fixed.\footnote{\url{https://bugs.chromium.org/p/project-zero/issues/detail?id=1930}}

\textbf{Autofill.}
Silver et al.~\cite{silver2014password} studied the autofill feature of ten password managers.
They demonstrated that if a password manager autofilled passwords without requiring user interaction, it was possible to steal a user's credentials for all websites that were vulnerable to a network injection attack or had an XSS vulnerability on any page of the website.
They also showed that even if user interaction was required, if autofill was allowed inside an iframe, then the attacker could leverage clickjacking to achieve user interaction without users realizing they were approving the release of their credentials.
Stock and Johns~\cite{stock2014protecting} also studied autofill related vulnerabilities in six browser-based password managers and had similar findings to Silver et al.

\textbf{Storage.}
Gasti and Rasmussen~\cite{gasti2012security} analyzed the security of the password vaults used by thirteen password managers, finding a range of vulnerabilities that could leak sensitive information to both passive and active attackers.
These vulnerabilities were related to unencrypted metadata as well as side channel information leakage from encrypted data.

Chatterjee et al.~\cite{chatterjee2015cracking} and Bojinov et al.~\cite{bojinov2010kamouflage} proposed alternative password vault schemes that are more resilient to offline attacks, but password managers have not adopted these schemes. 

A recent study by Independent Security Evaluators~\cite{underthehood} found that password managers were not encrypting passwords that they wrote to memory, making it trivial to extract some passwords from the password vault even when it was not in use.

\textbf{Usability.}
In 2006, Chiasson et al.~\cite{chiasson2006usability} conducted a usability study of two password managers, finding significant vulnerabilities due to users' incomplete mental models regarding how these password managers worked.
More recently, Fagan et al.~\cite{fagan2017investigation} surveyed users and non-users of password managers to better understand why people chose to adopt password managers.
They found that users adopted password managers primarily due to usability, not security benefits; in contrast, non-users generally avoid password managers due to security, not usability concerns.

Lyastani et al.~\cite{lyastani2018better} studied whether adoption of a password manager helped increase the strength of a user's passwords, finding that while users of password managers on average had stronger passwords than those of the general public, they still rarely had a unique, brute force-resistant password for every website.
Zhang et al.~\cite{zhang2019people} interviewed users to investigate how they use their password managers, finding that users of browser-based managers were more likely to reuse password than users of app-based or extensions-based password managers.

\textbf{Relation to This Work}
To our knowledge, our work is the first to study the strength of password generators in password managers and the first to simultaneously consider the full password manager lifecycle~\cite{choong2014cognitive} (i.e., generation, storage, and autofill).
Much of the work examining the security of password manager autofill and storage is now over five or more years old~\cite{silver2014password,stock2014protecting,li2014emperor,gasti2012security}.
As there have been significant updates to password managers in that time, we have replicated this early work to determine whether the password managers we studied have addressed the core issues revealed by this prior work, or whether they remain vulnerable.

\section{Analyzed Password Managers}
In this work, we analyzed 13 different password managers.
These password managers can be categorized based on their level of integration with the browser: app, extension, and browser.
We focused on password managers in the browser but included two desktop clients for comparison.
Apps are desktop clients that are not integrated with the browser.
Extension-based password managers are deployed as a browser extension and do not rely on a desktop application.
Browser-based password managers are native components implemented as part of the browser.
We chose from among the most popular systems within each of these categories.

\begin{table}
	\begin{adjustwidth}{-1.25cm}{}
		\setuptable
		
		\rowcolors{2}{gray!10}{}
		\begin{tabular}{l|l|ccccc|cccccc|}
			\headrow{} & \multicolumn{1}{l}{System} \headline	
			
			& \headrow{Supports generation} & \headrow{Supports autofill}
			& \headrow{Cloud sync for extension settings} & \headrow{Cloud sync for vault}
			& \headrow{CLI support} \headline
			
			& \headrow{Supports MFA} & \headrow{Lockable Vault} & \headrow{Login on separate tab or app} & \headrow{Has assessment tool} & \headrow{Clears clipboard}
			& \headrow{Open Source} \headline[1.7cm]
			\\ \hline
			
			\hc & KeePassX		
			& \full	& \none	& \none	& \none &	\none	& \none	& \full	& \full	& \none	& \full & \full \\
			
			\hc \headcol{-2}{App} & KeePassXC		
			& \full & \full	& \none	& \none &	\full	& \none	& \full	& \none	&	\none	& \full & \full	\\ \hline

			\hc & 1Password X	
			& \full & \full	& \none	& \full	&	\full	& \full & \full	& \none	& \full	& \none & \none	\\
			
			\hc & Bitwarden		
			& \full & \full	& \none	& \full	&	\full	& \full & \full	& \none	& \none	& \none & \full	\\
			
			\hc & Dashlane		
			& \full & \full	& \prt	& \full	&	\none	& \full & \full	& \none	& \full	& \none & \none	\\
			
			\hc & LastPass		
			& \full & \full	& \none	& \full	&	\full	& \full & \full	& \none	& \full	& \none & \none	\\
			
			\hc \headcol{-5}{Extension}
			& RoboForm		
			& \full & \full	& \none	& \full	&	\none	& \full & \full	& \none	& \full	& \none & \none	\\ \hline
			
			\hc & Chrome			
			& \prt & \full	& \na		& \full	& \none	& \full & \none	& \full	& \none	& \na		& \prt	\\
			
			\hc & Edge				
			& \none & \full	& \na		& \full	& \none	& \full & \none	&	\full	& \none	& \na		& \none	\\
			
			\hc & Firefox			
			& \none & \full	& \na		& \full	& \none	& \full & \none	& \full	& \none	& \na		& \full	\\
			
			\hc & IE 					
			& \none & \full	& \na		& \full	& \none	& \full & \none	&	\full	& \none	& \na		& \none	\\
			
			\hc & Opera 			
			& \none & \full	& \na		& \full	& \none	& \full & \none	&	\none	& \none	& \na		& \prt	\\
			
			\hc \headcol{-6}{Browser}
			& Safari 			
			& \prt & \full	& \na		& \full	& \none	& \full & \none	&	\full	& \none	& \na		& \prt	\\ \hline
		\end{tabular}
		
		\caption{Analyzed Password Managers}
		\label{tab:overview}
	\end{adjustwidth}
\end{table}

The breakdown of analyzed password managers into these categories is given in Table~\ref{tab:overview}.
This table also reports on features related to utility and usability---support for password generation and autofill, support for synchronizing extension settings and password vaults using the cloud, ability to use the password manager from a command line interface---as well as security---whether the tool supports multi-factor authentication (MFA), whether the password vault can be locked, whether the master password for the vault must be entered on its own tab or application (to prevent spoofing of this dialog~\cite{bravo2012operating}), whether the password manager provides a tool to assess the security of stored accounts and passwords, whether the manager clears passwords from the clipboard after they are copied, and whether the tool is open source.

In the remainder of this section, we discuss each password manager analyzed and indicate which version of the password manager we evaluated.
In-depth details regarding password generation, autofill, and storage are found in their respective sections.

\subsection{App}
The app-based password managers we analyzed eschew cloud syncing of vaults and settings in favor of manual synchronization to increase security.

\textbf{KeePassX (v2.0.3).}
KeePass is an app-based password manager originally built using the .NET platform and intended for use on Windows.
KeePassX is a cross-platform port of KeePass, replacing the .NET platform with the QT framework.

\textbf{KeePassXC (v2.3.4).}
KeePassXC is a fork of KeePassX intended to provide more frequent updates and additional features not found in KeePass or KeePassX (e.g., more options for password generation, a command line interface).
KeePassXC also provides a browser extension that interfaces with the app to autofill passwords in the browser.
In total, the KeePass family of applications is estimated to have 20 million users~\cite{underthehood}.

\subsection{Extension}
\anyone{Would be good to describe the UIs a bit here.}

Extensions lack permissions to clear the clipboard and so none of the extension-based password managers support this feature, leaving user passwords vulnerable to any application with clipboard access indefinitely. 
None of the extensions we analyzed supported synchronizing settings for the extension itself, requiring that users remember to correctly update these settings to match their security preferences for each new device they set up.
These extension settings include security critical options, such as whether to log out when the browser is closed, whether to use autofill, and whether to warn before filling insecure forms. 
The user experience for each of the extension-based password managers is mostly similar.

\textbf{1Password X (v1.14.1).}
1Password is estimated to have 15 million users~\cite{underthehood}.
1Password provides both an app-based client (1Password) and an extension-based client (1Password X); in this paper, we evaluated the extension-based client because it is the recommended tool if integration with the browser is desired (something we assume most users would want).\footnote{\url{https://support.1password.com/getting-started-1password-x/}}
While the security of both systems is similar, there are a few small differences---e.g., the password is cleared from the clipboard in the app, but not the extension.
Unique to 1Password, to initially download the password vault from the cloud it is necessary to enter a 128-bit secret key that was presented to the user when they generated their account, providing an extra layer of security to the cloud-based password vault.

\textbf{Bitwarden (v1.38.0).}
Bitwarden is unique within the extension-based password managers that we analyzed in that all of its functionality is available to non-paid accounts, whereas other password managers required a subscription to gain access to some features.

\textbf{Dashlane (v6.1908.3).}
Dashlane is estimated to have 10 million users~\cite{underthehood}.
In addition to storing the username and password for each website, Dashlane also tracks and synchronizes the following three settings on a per-site basis: ``always log me in'', ``always require [the master password]'', and ``Use [password] for this subdomain only.''
This feature provides a slight advantage when compared to other extension-based password managers that do not synchronize any extension settings.

\textbf{LastPass (v4.24.0).}
LastPass is estimated to have 16.5 million users~\cite{underthehood}, the most of any commercial password manager.

\textbf{RoboForm (v8.5.6.6).}
RoboForm is estimated to have 6 million users.\footnote{\url{https://www.roboform.com/business/features}}
Like 1Password, RoboForm offers both an app-based client and an extension-based client; in this paper, we evaluated the extension-based client for the same reason we took this approach with 1Password X.

\subsection{Browser}
Compared to both app-based and extension-based password managers, browser-based password managers lack many features.
While all browser-based password managers allow the cloud account storing the password vault to be protected using multi-factor authentication, none except Firefox enable this vault to be locked short of removing the account from the browser. Firefox provides the option to use a master password to restrict access to the password vault. 
As these password managers do not have settings to sync and never copy a password to the clipboard, those features are not applicable.

\textbf{Chrome (v71.0).}
Chrome has some support for generating passwords. It detects when a user might need a password and offers to generate a password for the user.
Unlike any other password manager, Chrome has basic functionality to try to detect the password policy.

\textbf{Edge (v42.17134).} \textbf{Firefox (v64.0).} \textbf{Internet Explorer (v11.523).} \textbf{Opera (v58.0.3135).}
These password managers are all similar in high-level functionality.

\textbf{Safari (v12.0).}
Safari can generate passwords when integrated with iCloud Keychain, though these passwords are always of the form ``xxx-xxx-xxx-xxx''.

\subsection{Updates for Password Managers}
Since we conducted our research, there have been some minor changes in several of the password managers: (1) KeePassXC has transitioned to using Argon2D as their default key derivation function, (2) LastPass has updated their password generation interface, removing the option to select the number of digits, and (3) RoboForm has updated their password generation interface, removing the option to select the number of digits and increasing the default password length to 16.
We are also aware of a couple more significant changes on the horizon: Firefox will transition to using Firefox Lockbox as its default password manager, and Edge will transition to being built on top of the Chromium project.

\section{Password Generation}
\label{sec:generation}
Password generation is the first step in the password manager lifecycle.
Of the 13 password managers in our evaluation, seven have full support for password generation---KeePassX, KeePassXC, 1Password X, Bitwarden, Dashlane, LastPass, and Roboform---and two have partial support---Chrome and Safari.
To provide a baseline by which to compare the password managers, we wrote a python script that generates passwords using \texttt{/dev/random} and the online Secure Password Generator\footnote{https://passwordsgenerator.net} (SPG), the first search result when searching for ``password generator'' on Google.

\begin{table*}[t]
	\setuptable
	
	\rowcolors{2}{gray!10}{}
	\begin{tabular}{l|l|cccccc|l|}
		\multicolumn{1}{l}{System} \headline 
		
		& \headrow{Abbreviation} \headline

		& \headrow{Supported lengths} & \headrow{Require diverse characters} & \headrow{Avoid difficult characters}
		& \headrow{Default length} & \headrow{Default composition}
		& \headrow{Preserve safe settings} \headline
		
		& \multicolumn{1}{c}{Symbol set}
		\\ \hline
		
		KeePassX		& kpx
		& 3--64		& \prt	& \full	& 16	&	ld		& \prt	& \verb|!"#$%&'()*+,-./:;<=>?@[\]^_`{}~|\verb~|~ \\
		
		KeePassXC		& kpxc                                
		& 1--128 	& \prt	& \full	& 16	&	ld		& \none	& \verb|!"#$%&'()*+,-./:;<=>?@[\]^_`{}~|\verb~|~ \\ \hline

		1Password X	& oneps
		& 8--50 	& \full	& \full	& 20	&	all		& \none & \verb|!#%)*+,-.:=>?@]^_}~| \\
		
		Bitwarden		& bw
		& 5--128 	& \full	& \full	& 14	&	ld		& \none & \verb|!#$%&*@^| \\
		
		Dashlane		& dlan
		& 4--28 	& \full	& \full	& 12	&	all		& \full & \verb|!"#$%&'()*+,-./:;<=>?@[\]^_`{}~|\verb~|~ \\
		
		LastPass		& lpass
		& 4--100 	& \prt	& \full	& 12	&	ld		& \none & \verb|!#$%&*@^| \\
		
		RoboForm		& robo
		& 1--99 	& \full	& \full	& 14	&	all	& \none 	& \verb|!#$%@^| \\ \hline
		
		Chrome			& chrm
		& $>1$ 		& \full	& \none	& 15	& all		& \na 	& \verb|!-.:_| \\
		
		Safari			& sfri
		& 15 			& \full	& \none	& 15	& all		& \na		& \verb|-| \\ \hline

		SPG					& psgn
		& 6--2048	& \full	& \full	& 16	& all		& \na		& \verb|!"#$%&'()*+,-./:;<=>?@[\]^_`{}~|\verb~|~ \\
		
		\texttt{/dev/random}			& dvrn
		& $>1$		& \none	& \none	& \na	& \na		& \na		& \verb|!"#$%&'()*+,-./:;<=>?@[\]^_`{}~|\verb~|~\texttt{\textvisiblespace}  \\ \hline
	\end{tabular}
	
	\caption{Overview of Password Generation Features}
	\label{tab:passgenoverview}
\end{table*}

\subsection{Settings and Features}
Table~\ref{tab:passgenoverview} provides a summary of configuration options, default settings, and features for each of the tools tested.
All password managers support ensuring that at least one character from each selected character set is included in the generated password, though this can be turned off in KeePassX, KeePassXC, and LastPass.
All password managers other than the browser-based password managers also have an option to avoid generating passwords that contain characters that may be difficult for users to read and/or memorize (e.g., hard to pronounce, looks similar to another character), though the exact characters removed are not consistent between password managers.

While all password managers support the same set of letters and digits ([A-Za-z0-9]), they each had different symbol sets.
KeePassXC had the largest symbol set, supporting all standard ASCII symbols (other than space) as well as supporting the extended ASCII symbol set.
KeePassX and Dashlane also support the standard ASCII symbols (other than space), but not the extended ASCII symbol set.
1Password supports just over half of the ASCII symbols (19 symbols), with the other systems supporting 8 or fewer symbols.
As expected, limiting the symbol set has a significant impact on the strength of generated passwords, the implications of which are discussed later in this paper.

One issue common in most password managers is that they save the last used settings as the new default settings.
While this might seem like a feature targeted at usability, it has the potential to cause users to use less than optimal settings when generating passwords.
In general, there are two reasons for users to change their password generation settings: (1) establishing safe default settings, (2) generating a password that conforms with a policy that is weaker than the default settings.
In the latter case, the newer, weaker settings will replace the older, stronger settings as the new defaults.
While users can manually restore their safer settings, there is no guarantee that they will do so.
Dashlane takes the optimal approach by not automatically saving the latest settings but giving the user the option to override the current defaults.
KeePassX takes a middle-of-the-road approach, saving the new settings for future passwords generated until the application is closed and opened again.

\subsection{Password Collection and Analysis}
\label{sub:generationmethodology}

\anyone{It would have been better to have done one of two things: use consistent settings for all tools. Use default settings for all tools. There is a mix of this right now.}
\anyone{We should have generated a set of passwords using all defaults, both length and settings.}

To evaluate the quality of passwords generated by the password managers, we first collected a large corpus of generated passwords from each password manager.
We use a variety of methods to generate passwords: existing command line interfaces (Bitwarden, our python tool), modifying the source code to add a command line interface (Chrome, KeePassX, KeeyPassXC), or using Selenium (1Password X, Dashlane, LastPass, RoboForm).
We were unable to analyze passwords for Safari as it does not have any mechanism for scripting password generation, though we did manually generate and analyze 100 passwords to check for any obvious problems and did not detect any.

Generation was parameterized by character classes---letters (l), letters and digits (ld), letters and symbols (ls), symbols and digits (sd), and all four classes together (all)---and password length---8, 12, and 20 characters long---in order to determine if these options had any effect on the randomness of generated passwords.
Most tools defaulted to requiring that generated passwords contain one character from each character set, with only Chrome, KeePassX, KeePassXC, and our python tool not having this option enabled.
For each password manager, character class, and password length we generated 1 million passwords, except 1Password X which does not allow passwords to be generated that only have symbols and digits.
This resulted in a corpus of 147 million passwords ($10\times5\times3-3$).

After collecting this data set, we analyzed its quality in terms of randomness and guessability.
There is no known way to prove that a pseudorandom generator is indistinguishable from random, so instead we leveraged a variety of analysis techniques, each attempting to find evidence of non-random behavior: Shannon entropy, $\chi^2$ test for randomness, the zxcbvn password analysis tool~\cite{wheeler2016zxcvbn}, and a recurrent neural net-based password guesser~\cite{melicher2016fast}.

Shannon entropy is used to check for abnormalities in the frequency of characters (not passwords) produced by each generator. 
The Shannon entropy of a set is a measure of the average minimum number of bits needed to encode a string of symbols based on the frequency of their occurrence. 
It is calculated as $-\sum_{i}^{} p_ilog_b(p_i)$.
While Shannon entropy is a bad measure for user-chosen passwords~\cite{bonneau2012science}, it is useful in evaluating the relative strength of random passwords.
Shannon entropy is not affected by the length of passwords, only by the number of distinct characters that can be present in a string and their relative frequency within the corpus.

The $\chi^2$ test for randomness is a simple statistical test for determining whether the difference between two distributions can be explained by random chance.
We used the $\chi^2$ test to evaluate each of our passwords sets independently and corrected our p-values using a Bonferonni correction\footnote{To represent this correction, all $p$ values are multiplied by 147, with a maximum value of 1.00.
For this reason, most p values reported are 1.00, as only clearly significant results stay significant with such a large correction.} to account for the multiple statistical tests from the same family.

The zxcbvn tool created by Daniel Wheeler~\cite{wheeler2016zxcvbn} is used to detect dictionary words and simple patterns that might be present in passwords, both potential examples of non-randomness.
zxcbvn also estimates the number of guesses a password cracker would take to break a password, which we use to understand if passwords are resilient to online and offline guessing.

In order to detect whether generated passwords had more subtle patterns than what zxcvbn could detect, we used the neural network password analyzer built by Melicher et al.~\cite{melicher2016fast}.
This analyzer uses a Long Short-Term Memory (LSTM) recurrent neural network (RNN) architecture to build a password guesser based on a training set.
As output, it produces a Monte Carlo estimation of how long it would take the trained password guesser to guess passwords in a test set.
The configuration files we used for training and testing are provided in Listing~\ref{lst:nnconfig} in Appendix~\ref{appx:generation}.
For each password corpus, we used 80\% of the passwords to train the neural network and tested against 20\% of the passwords.
Due to problems with the analyzer, we were only able to test passwords of length 8 and 12, as length 20 passwords would crash with an out of memory exception regardless of what settings were used.

While zxbcvn and the recurrent neural net are both used to evaluate the quality of randomness in the generated passwords, they also served to give approximations for how many guesses it would take for an online or offline guessing attack to try that password.
Passwords that require more than $10^6$ guesses are considered to be resilient against online attacks and passwords that require more than $10^{14}$ guesses are considered to be resilient against offline guessing~\cite{florencio2014administrator}.
Using this guess count, we were able to analyze whether the password managers were generating passwords that were vulnerable to these attacks.

\subsection{Results}

\paragraph{Password Strength:}

\begin{figure}[t]
	\centering
	
	\includegraphics[width=\columnwidth]{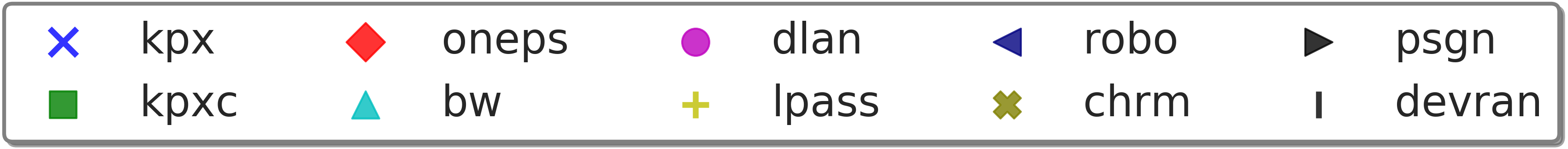}
	\hfill \vspace{-.5\baselineskip}
	
	\begin{subfigure}{.4\columnwidth}
		\includegraphics[width=\textwidth]{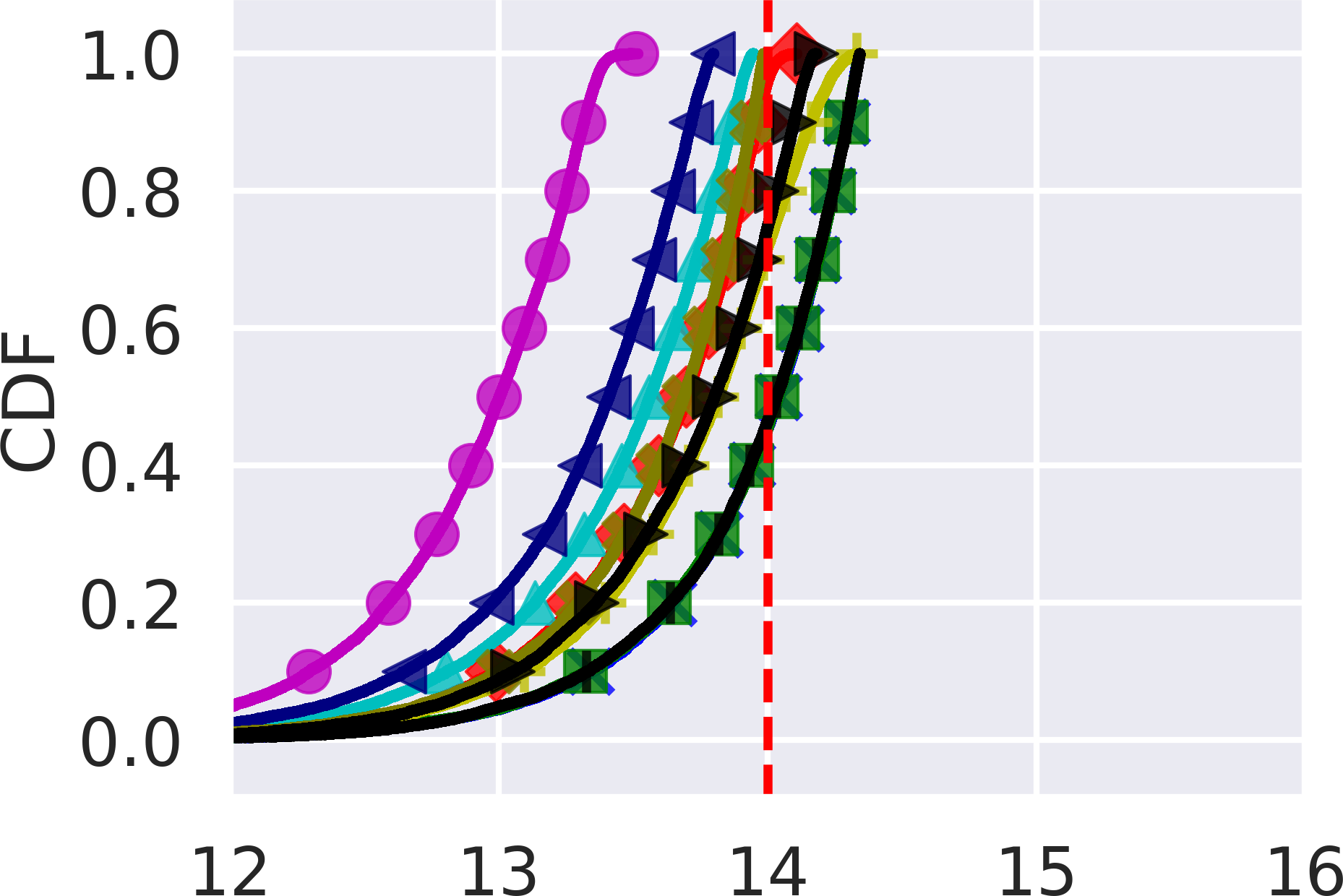}
		\caption{Length 8, ld}
		\label{fig:nn_8_ld}
	\end{subfigure}
	\begin{subfigure}{.4\columnwidth}
		\includegraphics[width=\textwidth]{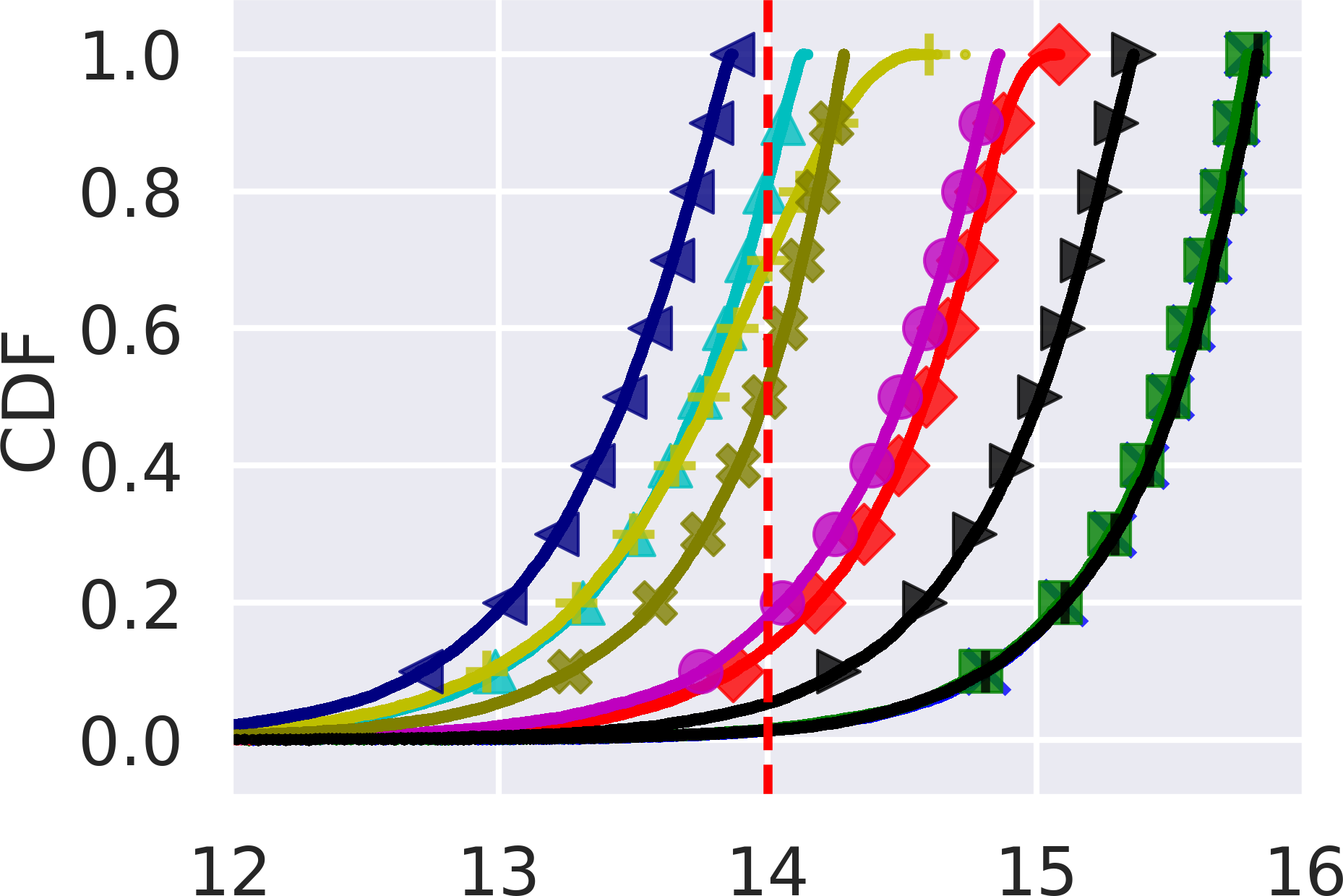}
		\caption{Length 8, all}
		\label{fig:nn_8_all}
	\end{subfigure}
	
	\begin{subfigure}{.4\columnwidth}
		\includegraphics[width=\textwidth]{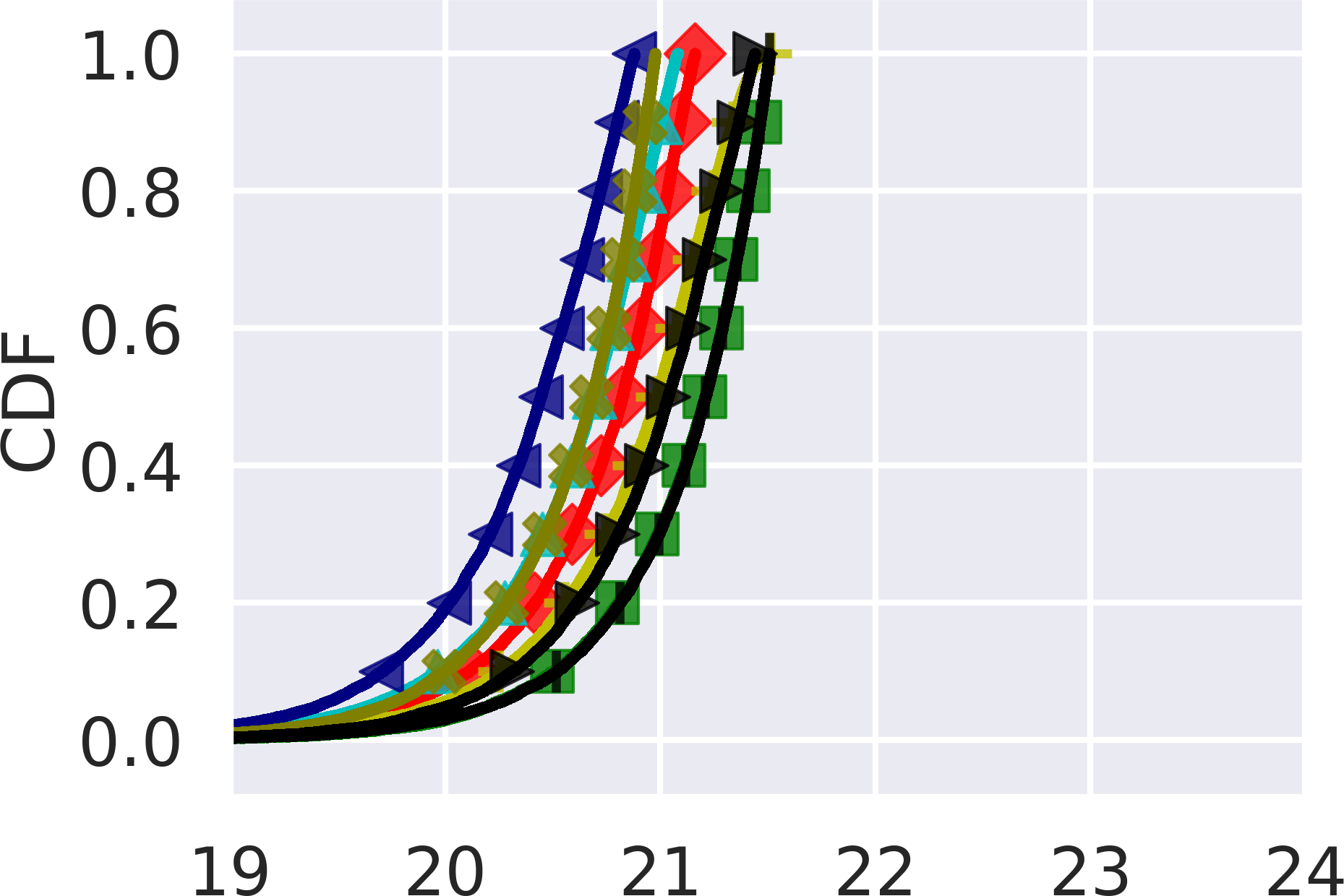}
		\caption{Length 12, ld}
		\label{fig:nn_12_ld}
	\end{subfigure}
	\begin{subfigure}{.4\columnwidth}
		\includegraphics[width=\textwidth]{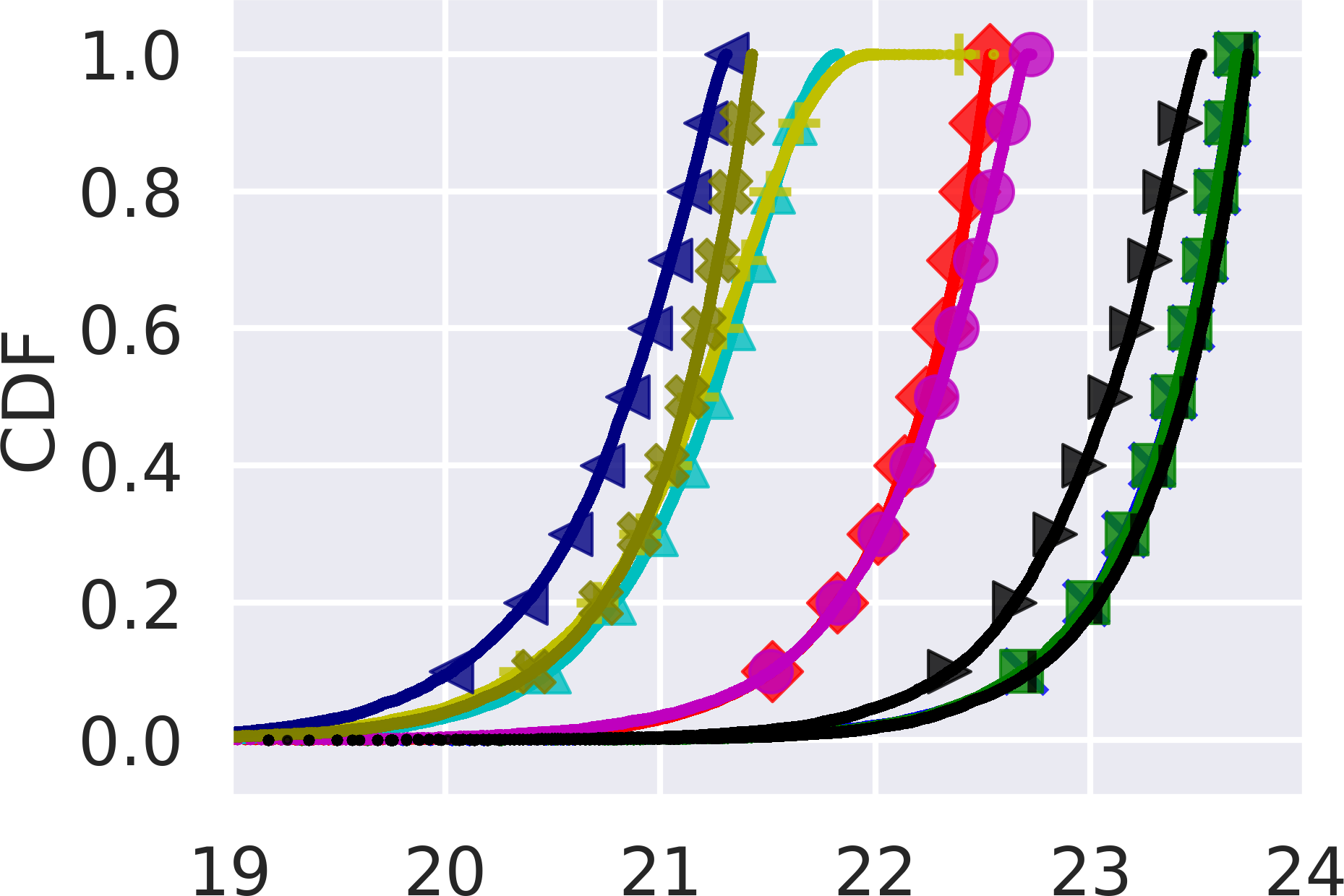}
		\caption{Length 12, all}
		\label{fig:nn_12_all}
	\end{subfigure}
	
	\label{fig:neuralnet}
	\caption{Neural Network Guess Estimates ($log_{10}$). Differences are primarily attributed to character set size.}
\end{figure}

\begin{table*}[t]
	\setuptable
	
	\rowcolors{2}{gray!10}{}
	\begin{tabular}{l|l|}
		\multicolumn{1}{l}{\adjustbox{rlap}{System}} & \multicolumn{1}{c}{Characters Sorted by Frequency} \\ \hline
		
		kpx   & \verb~'+,7lFr[AE/8"$OdNzGMn`_*3;D:i|~\verb|Z@s=#]whRb6~&Wm(2ck)\g^oy<aL}JCTq4e!->VI1BPvY9HSUjp{?5%xt0fX.uQK| \\
		kpxc  & \verb|NtpgT@vO<Be1hiY)H-`Kk;IXu^c4z$yqo6F/r>S_%Z3+U[=DL\as"0(2'VA?PdRm.:*jb]W~}Exn{f!Q|\verb~|7#CJw8G,&9lM5~ \\ \hline
		
		oneps & \verb|0314569782>^:*@.-~+%?,_V=a}N]!d)YjZK#ubeCATJUGBEDyozrgkMRtHwLvXWmqxfQhsniPFpc| \\
		bw		& \verb|%7#9532^@46!$&8*IYBomtbJFLUPVnXdzSexagHZrwusiMkpqcWNRvQKhfDCGAjyTE| \\
		dlan  & \verb|5473698QRHDNPAFMBKSCLYTXEGJijepnfgtryhbdkmqsxacz_*@~=){'[;&,!#.:"/$(^|\verb$|+}-%]?o`><\wuU2WvVZ$ \\
		dlan\textsuperscript{*}
					& \verb|3498576QNBHPAFMXJCYKTGSDRLEdqzmnpsfjghbtxckaioyre/$#{!<-,?"(\=].~*^'+`|\verb~|:;}>_)[%@&~ \\
		lpass	& \verb|%!$#&*@^jAGfRMOYPobszleTUiIwVhtDKNQqJgBSaWmpudcnLkEyHrZFxCXv3987542160| \\
		robo	& \verb|%#@!^$8624375HLYJXPDFCWAUENKVSTiRQGBMydgstkvqpfnjbwaemhrucox9Zz| \\ \hline
	
	chrm  & \verb|umSHDMeYNbnEGzCwaspZg6f:!XqLTBWrR9t5h3JP8Q7jc_iAFVK-kdxv2Uy.4| \\ \hline
	
	psgn  & \verb|4239750618LoQPYliRHpJkqIUZOnWBxmNhvdDbgAXtuVcwzysSCarMjEGKTfeF\!/(.+%}@|\verb+|'=[$`{?:*>&)~-;"^],<#_+ \\
	dvrn  & \verb|.\zdAP4L0^W,6@&+3w%?ebSqc-"Y$8EM'~QVu}iGojv(tK:y;I>#<TD_aU9C[lrH)/h5Z1|\texttt{\textvisiblespace}\verb~|sR`=mO]{*xXgnBNpfFJk2!7~ \\ \hline
	\end{tabular}
	
	\textsuperscript{*}Length 12 passwords. Dashlane uses different characters sets for long and short passwords.
	
	\caption{Character Frequencies for length 20 passwords using all characters. Groups of similar characters represent a requirement to include at least one character from that set, causing characters from smaller sets to be selected with greater frequency.}
	\label{tab:charfreq}
\end{table*}

Our analysis of the generated passwords found that nearly all passwords of length 12 and longer were sufficiently strong to withstand both online and offline guessing attacks (see Figures~\ref{fig:nn_12_ld} and \ref{fig:nn_12_all}).
Still, not all password managers created passwords of equal strength, with these small perturbations having a significant effect on the percentage of length 8 passwords that were secure against offline guessing attacks (nearly all were secure against online guessing attacks) (see Figures~\ref{fig:nn_8_ld} and \ref{fig:nn_8_all}).
These differences in strength can largely be explained by the different composition of character set classes used by each of the password managers.
While the difference is most pronounced when considering symbols (see Table~\ref{tab:passgenoverview}), several password managers also limit the available letters and digits (e.g., removing `0' and `O' due to similarity).
Looking at character frequencies (see Table~\ref{tab:charfreq}), we also found that Dashlane uses a different set of letters depending on the length of the passwords; it is unclear why Dashlane exhibits this behavior.

\paragraph{Randomness:}

\begin{table}[t]
	\setuptable
	
	\rowcolors{2}{gray!10}{}
	\begin{tabular}{l|lllll|}
		
		System & l & ld & ls & sd & all \\ \hline
		
		KeePassX		& \good & \good & \good & \good & \good	\\
		KeePassXC		& \good & \good & \good & \good & \good	\\ \hline
		
		1Password X & \bad & \good & \bad & \bad &				\\
		Bitwarden		& \bad & \good & \bad & \bad & \good	\\
		Dashlane		& \bad & \bad  & \bad & \bad & \bad	\\
		LastPass		& \bad & \good & \bad & \bad & \bad	\\
		RoboForm		& \bad & \bad  & \bad & \bad & \bad	\\ \hline
		
		Chrome			& \good	& \good & \good & \good & \good \\ \hline
		
		SPG									& \bad 	& \good 	& \bad 	& \bad 		& \bad \\
		\texttt{/dev/rand}	& \good & \good & \good & \good & \good \\ \hline
		
	\end{tabular}

	\rowcolors{3}{}{gray!10}
	\begin{tabular}{ll}
		\good & No statistically significant results (random)\\
		\bad  & Statistically significant result (non-random)	
	\end{tabular}

	\caption{$\chi^{2}$ test for random character distribution}
	\label{tab:chi2significance}
\end{table}

Our $\chi^2$ testing found several instances of non-random behavior in the generated passwords (see Table~\ref{tab:chi2significance}, detailed $\chi^2$ and $p$ values are in Tables~\ref{tab:chisquaredvalues8}--\ref{tab:chisquaredvalues20} in Appendix~\ref{appx:generation}).
All but one of the non-random character frequency distributions can be explained by a single feature---requiring that passwords have at least one character from each character set.
When this feature is not enabled, the probability that any given character will appear in a password is proportional to the length of the password, and the number of characters from all the enabled character sets (see Equation~\ref{eq:freq}).
When this feature is enabled, the probability is also proportional to the number of characters in that character set (see Equation~\ref{eq:freqadj}), causing character frequencies to be higher for characters that come from smaller character sets (e.g., digits, symbols), explaining the non-uniformity detected by the $\chi^2$ test.
We note that it would be possible to adjust for this skew and preserve a uniform distribution, though there no significant security effect from not correcting it.

\begin{equation}
\label{eq:freq}
length*\frac{1}{|characters_{all}|}
\end{equation}

\begin{equation}
\label{eq:freqadj}
((length-|sets|)*\frac{1}{|characters_{all}|})+\frac{1}{|characters_{set}|}
\end{equation}

While the results for Bitwarden (sd) and Dashlane (l) may at first not appear to follow this pattern, they in fact do.
Bitwarden (sd) has equal numbers of symbols and digits (see Table~\ref{tab:charfreq}, causing them to be selected with equal frequency.
In contrast, Dashlane (l) has a non-random distribution because it uses a different number of upper and lowercase letters.

The only non-random result that cannot be explained at least partially by this feature is RoboForm (l), which has an equal number of upper and lowercase characters.
Looking at all the character frequencies for RoboForm (see Table~\ref{tab:robocharfreq} in Appendix~\ref{appx:generation}) we find that uppercase letters, other than `Z', are selected more frequently than the lowercase letters.
Additionally, the characters `Z', `z', '9' are consistently the least frequently selected characters.
While it is not entirely clear what causes this issue, we hypothesize that it might be related to selecting characters using modular arithmetic (e.g., $rand()\%(max-min)+min$), which can have a slight bias to lower valued results. 

\paragraph{Random but Weak Passwords:}

\begin{table}[t]
	\setuptable
	
	\rowcolors{2}{gray!10}{}
	\begin{tabular}{l|lll|l|}
		\multicolumn{1}{c}{System} \headline[2cm] & \headrow{Length} & \headrow{Composition} & \headrow{Guesses ($log_{10}$)} \headline[2cm] & \multicolumn{1}{c}{Password} \\ \hline
		
		KeePassX		&	8		&	l		& 4.96 &	\verb|TaKEdeen|			\\
		KeePassXC		&	8		&	sd	& 4.84 &	\verb|'+'+'+_+|			\\ \hline
		
		1Password X	&	12	&	ls	& 8.76 &	\verb|oMMMMMMT?m*m|	\\
		Bitwarden		&	8		&	all	& 4.12 &	\verb|d@rKn3s5|			\\
		Dashlane		&	8		&	sd	& 4.48 &	\verb|////$8$8|			\\
		LastPass		&	12	&	all	& 8.92 &	\verb|B@KeRee22241|	\\
		RoboForm		& 8		& ls	& 5.02 &	\verb|SAWyE@rS|			\\
		RoboForm		&	8		&	sd	& 4.06 &	\verb|2345678#|			\\ \hline
		
		Chrome			& 8		& all	& 4.85 &	\verb|Tz5a5a5a|			\\ \hline
		
		SPG					& 8		&	ls	& 5.32 &	\verb|nW$nW$RR|			\\
		\texttt{/dev/rand} & 12 & l & 9.0 & \verb|MrKNxQNDAViS| \\ \hline
	\end{tabular}
	
	\caption{Randomly Generated Weak Passwords}
	\label{tab:weakpasswords}
\end{table}

In our analysis of the zxcbvn results, we found that occasionally all password managers would generate exceptionally weak passwords, examples of which are shown in Table~\ref{tab:weakpasswords}. 
While this is expected behavior for a truly random generator, it still results in suboptimal passwords.

Even though randomly generated length 8-character passwords have the potential to be resilient to offline attack (e.g., $log_{10}(96^{8}/2)=15.56$), password managers will present users with passwords of this length that are vulnerable to both online and offline attacks.
At length 12, the weakest passwords are no longer vulnerable to online attacks but are still vulnerable to offline attacks.
Finally, at length 20 the weakest passwords were able to withstand an offline attack.
While the occurrence of these weak passwords is relatively rare (less than 1 in 200), it is still preferable to choose passwords of sufficient length such that even randomly weak passwords are likely to be resilient to online and offline attacks.
Based on our analysis of these results, that is length 10 for resilience to online attacks and length 18 for resilience to offline attacks.

\section{Password Storage}\label{sec:storage}
Password storage is the second stage of the password manager lifecycle.
To evaluate the security of password storage, we manually examined the local password databases created by each password manager, looking to see what information was and was not encrypted, as well as examining how changes in the master password effected the encryption of data.
We determined how encryption took place through a combination of claims from the password manager's maintainer, options available in the client, and format of the ciphertext.
We focus on the storage of the password vault on the local system as the cloud databases are not available to us for direct evaluation.
An overview of this information is provided in Table~\ref{tab:storageoverview}.

\begin{table*}
	\setuptable
	
	\rowcolors{4}{gray!10}{}
	\begin{tabular}{l|l|llrc|ccc|cccc|cc|}
		\multicolumn{2}{l}{} \headline

		& \headrow{Encryption} & \headrow{KDF} & \headrow{KDF Rounds} & \headrow{Requires strong MP} \headline

		& \headrow{URL} & \headrow{Icon} & \headrow{Username} \headline

		& \headrow{Creation time} & \headrow{Modification time} & \headrow{Last use time} & \headrow{Fill count} \headline
		
		& \headrow{User's email} & \headrow{User's settings} \headline
		\\ \cline{3-15}
		
		System & Storage & \multicolumn{4}{c|}{Storage Encryption} & \multicolumn{9}{c|}{Metadata Encrypted} \\
		\hline
		
		KeePassX		& File (.kbdx)		& AES-256		& AES-KDF	& 100,000	&\none	
		&\full		&\full		&\full		&\full		&\full		&\full		&\full		&\full		&\full		\\
		KeePassXC		&	File (.kbdx)		& AES-256		& AES-KDF	& 100,000	&\none	
		&\full		&\full		&\full		&\full		&\full		&\full		&\full		&\full		&\full		\\ \hline
		
		1Password X	&	File (.json)		& AES-256		& PBKDF2	& 100,000	&\prt
		&\full		&\full		&\full		&\full		&\full		&\full		&\full		&\none		&\none		\\
		Bitwarden		&	File (.json)		& AES-256		& PBKDF2	& 100,001	&\prt		
		&\full		&\full		&\full		&\full		&\none		&\full		&\full		&\none		&\full		\\
		Dashlane		&	File (.aes)			& AES-256		& Argon2D	& 3				&\prt		
		&\full		&\none		&\full		&\full		&\full		&\full		&\full		&\none		&\full		\\
		LastPass		&	File (.sqlite)	& AES-256		& PBKDF2	& 100,100	&\prt	
		&\full		&\full		&\full		&\full		&\full		&\full		&\full		&\none			&\full		\\
		RoboForm		&	File (.rfo)			& AES-256		& PBKDF2	& 4,096		&\prt		
		&\full		&\full		&\full		&\full		&\full		&\full		&\full		&\none		&\full		\\ \hline
		
		Chrome			&	File (.sqlite)\textsuperscript{1}	& OS				&\na			&\na			&\na		
		&\none		&\na			&\none		&\none		&\full		&\none		&\none		&\full		&\full		\\
		Edge				&	Windows Vault		& \na				&\na			&\na			&\na		
		&\full		&\na			&\full		&\full		&\full		&\full		&\full		&\full		&\full		\\
		Firefox			&	File (.json)		& 3DES			&SHA-1			&1			&\none	
		&\none		&\na			&\full		&\none		&\full		&\none		&\none		&\full		&\full		\\
		IE					&	Windows Vault		& \na				&\na			&\na			&\na		
		&\full		&\na			&\full		&\full		&\full		&\full		&\full		&\full		&\full		\\
		Opera				&	File (.sqlite)\textsuperscript{1}	& OS				&\na			&\na			&\na		
		&\none		&\na			&\none		&\none		&\full		&\none		&\none		&\full		&\full		\\
		Safari			&	OSX Keychain		& \na				&\na			&\na			&\na		
		&\full		&\na			&\full		&\full		&\full		&\full		&\full		&\full		&\full		\\ \hline
	\end{tabular}

	\rowcolors{1}{}{}
	\begin{tabular}{l}
		\textsuperscript{1}On Linux, Chromium-based browser attempt to store the password in the GNOME keyring or KWallet 4.\\
		If neither of these are available, it will store the passwords in plaintext~\cite{chromiumLinux}.
	\end{tabular}

	\caption{Overview of Password Vault Encryption}
	\label{tab:storageoverview}
\end{table*}

\anyone{We should have looked at storage on mobile for the browser-based managers.}

\subsection{Password Vault Encryption}
The app-based and extension-based password managers all encrypt their databases using AES-256.
These systems all use a key derivation function (KDF) to transform the master password (MP) into a cryptographic key that can be used for encryption.
KeePassX and KeePassXC use AES-KDF with 100,000 rounds.
All of the extension-based password managers, other than Dashlane, use PBKDF2, with only RoboForm using less than 100,000 rounds.
Dashlane is the only password manager that uses a memory-hard KDF, Argon2D, with 3 rounds.
While not used by default, KeePassXC does support the option of using Argon2D in place of PBKDF2.

Each of these password managers has different requirements for the composition of the master password.
KeePass and KeePassX both allow any composition for the master password, including not using a master password at all.
The extension-based password managers all require a master password but vary in composition requirements.
LastPass, RoboForm, and Bitwarden require that the master password be at least eight characters but impose no other restrictions.
1Password X increases the minimum length to 10, but otherwise is the same as the other three.
Only Dashlane has compositions requirements, requiring a minimum length of 8 characters and one character from each character class (lowercase, uppercase, digit, symbol).

Of the browser-based password managers, only Firefox handles the encryption of its password vault itself.
It uses 3DES to encrypt the password data, using a single round of SHA-1 to derive the encryption key.
It imposes no policy on the master password.
Compared to other password managers that handle their own encryption, Firefox is by far the weakest.

The remaining browser-based systems rely on the operating system to help them encrypt the password vault.
Edge, Internet Explorer, and Safari all rely on the operating systems keyring to store credentials.
For Edge and Internet Explorer this is the Windows Vault; for Safari it uses the macOS keychain.

Chrome and Opera also rely on the operating system to encrypt the password, but how they do so varies by operating system.
On Windows, the \texttt{CryptProtectData} function is used to have Windows encrypt the password with a key tied to the current user account.
On Linux, these systems first try to write the password to the GNOME keyring or KWallet 4, falling back to storing the passwords in plaintext if neither of these keychains is available.
On macOS, the passwords are encrypted with keys derived by the macOS keychain, though the website passwords themselves are stored locally rather than on the keychain.

Browser-based password managers, other than Firefox, rely on the operating system to encrypt passwords and therefore do not allow users to establish a master password.
As such, there is no way to lock the password vault separately from locking the account.
While outside the scope of this paper, we also note that there is a need for more research examining the security of OS-provided encryption functions and keychains.

\subsection{Metadata Privacy}
Compared to earlier findings by Gasti and Rasmussen~\cite{gasti2012security}, we find that app-based and extension-based password managers are much improved in ensuring that metadata is properly protected.
KeePassX and KeePassXC both encrypt all metadata.
Extension-based password managers encrypt most metadata, but all have at least one item they do not.

1Password X stores extension settings in plaintext, allowing them to be read or modified by an attacker.
These settings include security-related settings such as whether auto-lock is enabled, default password generation settings, and whether to show notifications.
While Dashlane encrypts the website URLs, it does not encrypt the website icons it associates with those URLs, allowing an attacker to infer websites for which a user has accounts. 
All extension-based password managers leak the email address used to log in to the password manager.

Browser-based managers that rely on an operating system provided keychain (Edge, Internet Explorer, Safari, as well as Chrome and Opera on Linux) use these tools to protect all relevant metadata.
For the other browser-based password managers (Chrome and Opera on Windows and macOS, as well as Firefox on all operating systems), there is a significant amount of unencrypted metadata.
All three of these password managers store the URL in cleartext, and only Firefox encrypts the username.
They also reveal information about when the account was created, when it was last used, and how many times the password has been filled.

%

\newcommand*{\low}{\LEFTCIRCLE}

\section{Password Autofill}\label{sec:autofill}

\begin{table*}
	\setuptable
	
	\rowcolors{4}{gray!10}{}
	\begin{tabular}{l|ccc|cc|ccc|cccc|cc|cc|}
		\multicolumn{1}{l}{} \headline[4cm] &
		\headrow{Interaction Required for HTTPS} & \headrow{Interaction Required for bad cert} & \headrow{Interaction Required for HTTP} \headline[4cm] &
		\headrow{Won't fill same-origin iframe} & \headrow{Won't fill cross-origin iframe} \headline[4cm] &
		\headrow{Won't fill different URL} & \headrow{Won't fill HTTPS$\rightarrow$bad cert} & \headrow{Won't fill HTTPS$\rightarrow$HTTP}  \headline[4cm] & \headrow{Won't fill different \texttt{action} (static)} & \headrow{Won't fill different \texttt{action} (dynamic)} & \headrow{Won't fill different \texttt{method}} & \headrow{Won't autofill different \texttt{input} fields} \headline[4cm] &
		\headrow{Won't fill \texttt{type=``text''} field} & \headrow{Won't fill non-login form fields} \headline[4cm] &
		\headrow{Fills password on transmission} & \headrow{Obeys \texttt{autocomplete=``off''}} \headline[4cm]
		\\ \cline{2-17}
		
		System &
		\multicolumn{3}{c|}{Interaction} &
		\multicolumn{2}{c|}{iframe} &
		\multicolumn{7}{c|}{Difference in fill form} &
		\multicolumn{2}{c|}{Fields} &
		\multicolumn{2}{c|}{Misc}
		\\ \hline
		
		KeePassXC
		&\prt		&\prt		&\prt		
		&\prt		&\none	
		&\none	&\none	&\full	&\full	&\full	&\none	&\none
		&\full	&\full
		&\none	&\none
		\\ \hline
		
		1Password X
		&\full	&\full	&\full		
		&\prt		&\full	
		&\prt		&\prt		&\full	&\prt		&\prt		&\prt		&\prt
		&\full	&\full
		&\none	&\none
		\\
		
		Bitwarden
		&\prt		&\prt		&\prt		
		&\prt		&\full	
		&\none	&\none	&\none	&\none	&\none	&\none	&\none
		&\full	&\none
		&\none	&\none
		\\
		
		Dashlane
		&\none	&\none	&\full		
		&\prt		&\full	
		&\none	&\none	&\full	&\none	&\none	&\none	&\none
		&\none	&\full
		&\none	&\none
		\\
		
		LastPass
		&\none	&\none	&\full		
		&\prt		&\full	
		&\none	&\none	&\prt		&\full	&\none	&\none	&\prt
		&\full	&\full
		&\none	&\none
		\\
		
		RoboForm
		&\prt		&\prt		&\prt		
		&\prt		&\full	
		&\none	&\none	&\none	&\none	&\none	&\none	&\none
		&\full	&\full
		&\none	&\none
		\\ \hline
		
		Chrome
		&\none	&\full	&\none		
		&\none	&\prt	
		&\none	&\full	&\full	&\none	&\none	&\none	&\none
		&\full	&\none
		&\none	&\none
		\\
		
		Edge
		&\none	&\full	&\full		
		&\none	&\prt	
		&\none	&\prt		&\full	&\none	&\none	&\none	&\none
		&\full	&\none
		&\none	&\none
		\\
		
		Firefox
		&\none	&\none	&\none		
		&\none	&\none	
		&\none	&\none	&\full	&\full	&\full	&\none	&\none
		&\full	&\none
		&\none	&\none
		\\
		
		IE
		&\none	&\full	&\full		
		&\none	&\prt	
		&\none	&\prt		&\full	&\none	&\none	&\none	&\none
		&\full	&\none
		&\none	&\none
		\\
		
		Opera
		&\none	&\full	&\none		
		&\none	&\prt	
		&\none	&\full	&\full	&\none	&\none	&\none	&\none
		&\full	&\none
		&\none	&\none
		\\
		
		Safari
		&\full	&\full	&\full		
		&\prt		&\prt	
		&\prt		&\prt		&\prt		&\prt		&\prt		&\prt		&\prt
		&\full	&\full
		&\none	&\none
		\\ \hline

	\end{tabular}
	
	\caption{Overview of Password Autofill Features}
	\label{tab:autofilloverview}
\end{table*}

Of the password managers we evaluated, only KeePassX did not support autofill in the browser\footnote{There is a browser extension adding autofill for KeePassX, but it is a third-party tool not a part of the KeePassX project.} and Bitwarden warns that its autofill functionality is experimental.
To evaluate these tools, we developed websites that leveraged the attacks identified by Li et al.~\cite{li2014emperor}, Silver et al.~\cite{silver2014password}, and Stock and Johns~\cite{stock2014protecting}.
We also updated these attacks to address protections that have been added by browsers and password managers since the attacks were first described.
Table~\ref{tab:autofilloverview} highlights several of our findings.

\subsection{User Interaction Requirements}
If an attacker can compromise a web page using either a network injection or XSS attack, they can insert malicious JavaScript that will steal the user's password when it is entered.
If a password manager autofills passwords without first prompting the user, then the user's password will be surreptitiously stolen simply by visiting the compromised website.
As such, user interaction should ideally be required before autofill occurs.
Of the password managers we tested, only 1Password X and Safari always require user interaction before filling in credentials.
The remaining password managers exhibited different behavior depending on the protocol the website was served over (i.e., HTTPS or HTTP) as well as whether the HTTPS certificate was valid.

For websites served over HTTPS with a valid certificate, KeePassXC, Bitwarden, and RoboForm require user interaction by default, but also allow user interaction to be disabled.
Dashlane, Lastpass, and Firefox default to autofilling passwords without user interaction, though there is an option to require user interaction.
Chrome, Edge, Internet Explorer, and Opera always autofill user credentials.
While having an option to require user interaction (Dashlane, LastPass, Firefox) is preferable to lacking that option (Chrome, Edge, Internet Explorer, Opera), in practice the results are likely the same for most users (who are unlikely to change their default options).

While network injection attacks are still possible on sites using HTTPS (i.e., TLS man-in-the-middle attacks~\cite{oneill2016tls}), they are much easier to accomplish and more likely if the HTTPS certificate is invalid.
Reasons for a bad HTTPS certificate range from benign (e.g., expired by a day) to malicious (e.g., invalid signature, revoked).
In both cases, password managers should altogether reject filling in the password or at the least require user interaction before autofilling the password.
In the case of an invalid certificate, KeePassXC, Bitwarden, RoboForm, Dashlane, Lastpass, Firefox all function as they did with a valid certificate.
Edge and Internet Explorer both change their behavior and always require user interaction for bad certificates.
Chrome and Opera also change their behavior, entirely disabling the ability to autofill passwords.

Network injection attacks are also more likely and easier to accomplish when the website is served using an unsecured connection (i.e., HTTP).
As with bad certificates, password managers should refuse to autofill the password or require user interaction before filling it in. 
KeePassXC, Bitwarden, and RoboForm continue to require user interaction by default, but do allow users to disable this requirement.
Dashlane, LastPass, Edge, and Internet Explorer all change their behavior to always require user interaction before autofilling passwords on HTTP websites.


%
%

\subsection{Autofill for iframes}
Autofilling passwords within iframes is especially dangerous, regardless of whether user interaction is required or not~\cite{silver2014password,stock2014protecting}.
For example, clickjacking can be used to trick users into providing the necessary user interaction to autofill their passwords, allowing an attacker to steal passwords for vulnerable websites loaded in an iframe (same-origin or cross-origin).
Even worse, if autofill is allowed for cross-domain iframes and user interaction is not required, then the attacker can programmatically harvest the user's credentials for all websites where the attacker can perform a network injection or XSS attack (by loading compromised websites into iframes).

For both the clickjacking and harvesting attacks, the user must first visit a malicious website which will then launch the attacks, but this is often not a significant obstacle for an adversary.
In the worst case, if a system is vulnerable to a harvesting attack and the attacker has access to the user's WiFi access point (e.g., at hotel or airport)---allowing them to trivially conduct network injection attacks---then all of a user's credentials can surreptitiously be stolen when the user views the network login page for the compromised access point~\cite{silver2014password,stock2014protecting}

KeePassXC, 1Password X, Dashlane, and LastPass autofill within same-origin iframes, leaving them vulnerable to clickjacking attacks.
Bitwarden and RoboForm also autofill within same-origin iframes, though if user interaction is required they are largely immune to clickjacking as this interaction happens outside of the website inside the extension drop-down.
All of the browsers will autofill within a same-origin iframe.

KeePassXC does allow autofill for cross-domain iframes; while by default it does require user interaction before autofill in cross-domain iframes, this requirement can be disabled leaving KeePassXC vulnerable to the harvesting attack described above.
Of the extension-based password managers, 1Password X, LastPass, and RoboForm will not fill autofill within a cross-origin iframe.
Bitwarden and Dashlane do autofill cross-origin iframe, but autofill the password for the domain of the top-most window (i.e., domain displayed in the URL bar), preventing an attacker from stealing the cross-domain credentials.

Chrome, Edge, Internet Explorer, Opera, and Safari all require user interaction before they will autofill passwords into a cross-domain iframe, though this still leaves them vulnerable to clickjacking attacks.
Firefox defaults to not requiring user interaction before autofilling passwords into cross-domain iframes, leaving it vulnerable to the domain harvesting attack by default.



\subsection{Fill Form Differing from Saved Form}
Password managers detect when a user manually enters a password into a login form and will then offer to save that password for later use.
When the password manager later fills this password, it can check that the form to be filled is similar to the form used when the password was saved (e.g., same path or protocol).
These types of checks help ensure that the user is entering their password in a non-compromised form that has security equivalent to the form they were using when they first saved their password.
Still, there are many situations where it makes sense for the form to have changed---for example, the password was saved on a registration form. (i.e., not a login form).

As such, we gave password managers a full-dot if they either disallowed filling the form or showed the user a notification when there was some disparity between the fill form and the form used to save the password.
A half-dot was given if the password manager required user interaction when there was a disparity, but only if this user interaction couldn't be disabled (as it can be in Bitwarden and RoboForm).
Note that 1Password X and Safari always require user interaction and therefore always receive at least a half-dot.
In the results discussed below, we only highlight when password managers act differently due to discrepancies in the login form.

Password managers do not react to discrepancies in the URL the form is served at (other than checking that the domains match).
If the password was saved on a form served over HTTPS, Chrome and Opera will refuse to fill it in a form served with a bad HTTPS certificate, with Edge and IE requiring user interaction.
If the form is instead served over HTTP, 1Password X and Dashlane will warn users and Chrome, Edge, Firefox, IE, and Opera will refuse to fill the password.
Also, LastPass will force user interaction.

If when the page is first loaded there is discrepancy in the form's \texttt{action} property (the URL the password will be submitted to), KeePassXC, LastPass, and Firefox will display a warning, with Firefox also refusing to fill the password.
If the \texttt{action} property is changed after page load (i.e., dynamically), KeePassXC and Firefox will display a warning, though unlike before Firefox will go ahead and fill the password.
Passwords managers do not react to a similar discrepancy in the \texttt{method} property.
If the \texttt{input} fields in the form have been renamed or removed, LastPass will require user interaction.

\subsection{Non-Standard Login Fields}
We investigated whether password managers would fill form fields with \texttt{type=``text''} (as opposed to \texttt{type=``password''}), finding that only DashLane would autofill the password in this case.
We also examined whether the tools would autofill a minimal form (i.e., a non-login form), containing only two input fields: a text field and a password field; autofilling in this situation reduces the effort required for an attacker to harvest credentials.
In this case, we found that Bitwarden, Chrome, Edge, Firefox, IE, and Opera would all autofill these non-login forms, with the remaining browsers only filling them when explicitly requested to by the user.

\subsection{Potential Mitigation}
\label{sec:mitigation}
Stock et al.~\cite{stock2014protecting} recommended a more secure form of autofill that would address XSS-vulnerabilities.
Instead of filling the password onto the webpage, where it would be vulnerable to XSS attacks, a nonce was filled into the website as the password.
When the nonce was about to be transmitted on the wire to the website, the password manager would then replace the nonce with the real password.
This approach prevents JavaScript on the webpage from ever being able to access the user's password.
Additionally, the password manager can check that the password is being sent only to the website associated with the password and that the password form is not submitting to a different website.

We checked all the password managers to see if they supported this functionality and found that none of them did.
In our investigation of this feature, we tried to implement it ourselves and found that browsers did not allow extensions to modify the request body, preventing extension-based password managers from leveraging this more secure mode of operation.\footnote{It may be possible to allow extensions to support this functionality in Internet Explorer using its COM-based extensions, though the documentation is unclear in this regard.}
Enabling secure password entry is an area where browsers could do more to improve authentication on the web and is discussed in greater depth in Section~\ref{sec:discussion}.

Silver et al.~\cite{silver2014password} and Stock and Johns~\cite{stock2014protecting} also explored whether setting the \texttt{autocomplete} attribute to ``off'' on the password field would prevent password managers from storing or autofilling the password.
We found that no password manager obeys this attribute.

Looking at the current W3C specification, it is unclear whether the \texttt{autocomplete} attribute should preclude storage and autofill of login credentials~\cite{w3forms}.
While the specification does state that the ``user agent'' should not fill fields marked with \texttt{autocomplete}, it is unclear if this is only referring to primary user agent (i.e., the browser) or also user agent extensions (i.e., the password manager).
Mozilla's documentation also notes that in order to support password manager functionality, most modern browsers have explicitly chosen to ignore the autocomplete attribute for login fields.~\cite{mozAutocomplete}.
This helps explains why no password managers currently obey this parameter, even though in prior research there was some support for this attribute in browsers~\cite{silver2014password,stock2014protecting}.

\subsection{Web Vault Security \& Bookmarklets}
In their analysis of extension-based password managers, Li et al.~\cite{li2014emperor} showed that problems with the security of online password vaults could magnify autofill issues.
These web vaults include both standalone interfaces to the password vault as well as acting as the synchronization backend for extension-based password managers.
For example, cross-site request forgery (CSRF) could be used to change the URL associated with a set of credentials, allowing all the user's credentials to be autofilled and  stolen from a single malicious domain.
Alternatively, XSS vulnerabilities on a web vault could be used to steal all its passwords.

We evaluated the five extension-based password managers and their web vault backends to see if they had properly addressed potential CSRF and XSS attacks.
We found that 1Password X, Bitwarden, DashLane, and LastPass use CSRF tokens to prevent CSRF attacks.
RoboForm does not appear to use CSRF tokens and we were able to launch a CSRF attack against its web vault that changed the session timeout parameter.
We were unable to find other CSRF attacks as the web vault appears to use cryptographic authentication and not cookies to authenticate other requests.

To evaluate the susceptibility of the web vaults to XSS attacks, we manually inspected each web vault's content security policy (CSP) headers.
%
The results of this evaluation found no issues with either 1Password X or Dashlane's CSP policies.
Bitwarden's policies had two small issues: \texttt{script-src} allows ``self'' and \texttt{object-src} allows ``self'' and ``blob:''.
LastPass's policies allow for ``unsafe-inline'' in the \texttt{script-src}, leaving a significant opening for XSS attacks.
RoboForm did not have any CSP policy for their website.
We did try to craft XSS exploits for both LastPass and RoboForm, but these efforts were unsuccessful as both sites employed extensive input sanitization; regardless, both web vaults would benefit from implementing stricter (or any) CSP policies.

Finally, we examined whether extension-based password managers still have bookmarklet-based deployment options (used to support mobile devices) that are vulnerable to attack~\cite{li2014emperor}.
We found that other than LastPass, the extension-based password managers no longer support a bookmarklet-based deployment.
In their place, password managers rely on native mobile applications to handle password management on mobile devices.
LastPass's bookmarklets correctly execute code inside a protected iframe and filter dangerous messages sent to the bookmarklet, addressing the types of problems found by Li et al.~\cite{li2014emperor}.

\anyone{We should have looked at bug bounty programs.}

%

\section{Discussion}\label{sec:discussion}
Our research demonstrates that app-based and extension-based password managers are improved compared to how these types of tools performed in prior studies~\cite{gasti2012security,li2014emperor,silver2014password,stock2014protecting}.
In general, they have done a good job at addressing specific vulnerabilities: improving the protection of metadata stored in password vaults, removed (insecure) bookmarklets, limited the ability to autofill in iframes (preventing password harvesting attacks), and addressed web security problems in the online password vaults.
On the other hand, there has been little change from earlier work in how they handle passwords for areas without specific vulnerabilities: warning users about discrepancies between the fill form and form where the password was saved or implementation of XSS mitigations.
Similarly, browsers-based password managers continue to significantly lag behind app-based and extension-based password managers, both in terms of security and functionality.

%
%

Based on our findings, we recommend that users avoid Firefox's built-in password manager.
In particular, its autofill functionality is extremely insecure, and it is vulnerable to a password harvesting attack~\cite{silver2014password,stock2014protecting}.
If an attacker can mount network injection attacks against a user (e.g., control a WiFi access point), then it is trivial for that attacker to steal all credentials stored in the user's Firefox password vault.
Hopefully, these issues will be addressed when Firefox transitions to their Firefox Lockbox password manager.
Users of KeePassXC's browser extension should also ensure that they do not disable the user interaction requirement before autofill, as doing so will also make the client susceptible to the same password harvesting attack.

We also suggest that users should eschew browser-based password managers in favor of app- and extension-based password managers, as the latter are generally more feature rich, store passwords more securely, and refuse to fill in passwords in a cross-origin iframe.
The one exception to this is Safari's password manager, which does a good job of storing passwords and avoids autofill mistakes, though it does lack a good password generator.

With the app- and extension-based password managers there is still a need for users to ensure that they are properly configured.
Neither Dashlane nor LastPass require user interaction before autofilling passwords into websites, and Bitwarden and Roboform allow this interaction to be disabled.
If user interaction is disabled, a user that visits a compromised website (e.g., an attacker has exploited an XSS vulnerability) can have their password for that site stolen without the user being aware that this has happened.
While this is not as bad as a password harvesting attack~\cite{silver2014password,stock2014protecting} (which is now prevented by extension-based password managers), it is still a vulnerability that users should not need to know or worry about.
Of the extension-based password managers we studied, only 1Password X refuses to ever autofill passwords.

In the remainder of this section, we describe our recommendations to improve functionality within existing password managers.
We also identify several areas for future research that have the potential to significantly improve the utility and security of password managers.

\subsection{Recommendations}

\textbf{Filter weak passwords.}
Our research shows that password managers will randomly generate passwords that can be trivially cracked by online- or offline-guessing attacks.
This is a natural extension of password generation being truly random---i.e., any password can be generated, even if it is a natural language word with common substitutions (e.g., ``\verb|d@rKn3s5|'') or exhibits repeated characters patterns (e.g., ``\verb|'+'+'+_+|'').
While this is extremely unlikely for passwords of sufficient length (10 characters for online resistance, 18 for offline resistance), it is still possible.
To address this problem, we recommend that password generators add a simple filter that checks if the generated password is easily guessable (easily checked using zxcvbn), and if so, generate a replacement password.

\textbf{Better master password policies.}
Password managers require that users select and manage a master password, with the hope because they only need one password that users will select a sufficiently strong secret.
If users fail to pick a good master password, especially if the selected master password is not online-attack resilient, then a password manager becomes a single point of failure for that user's accounts.
Unfortunately, trusting users to always choose strong master passwords is problematic for three reasons: (1) users don't necessarily understand what constitutes a strong password, (2) their chosen passwords might have transformations they consider unique but turn out to be common, and (3) users might still select an easy password because it is more convenient.

For these reasons, we recommend that password managers adopt stringent requirements for master password selection, preventing users from turning their password manager into a single point of failure.
Additionally, password managers should all transition to using memory hard KDFs for transforming the master password into an encryption key. 

\textbf{Safer autofill.}
Autofilling credentials without user interaction puts those credentials at risk if the website is compromised by an XSS attack.
For this reason, we recommend that password managers default to require user interaction before autofilling passwords.
Where possible, we also suggest removing the option to disable user interaction as users are unlikely to understand the implications of turning it off.
Autofilling into iframes, same- or cross-origin, is also dangerous as it allows clickjacking attacks to circumvent user interaction requirements.
As such, we recommend disabling autofill with iframes, or if that is not feasible to consider moving the user interaction out of the web page and into the browser---as Bitwarden and RoboForm do---making clickjacking attacks much more difficult.

\subsection{Future Work}

\textbf{Browser-Supported Password Managers.}
Currently, authentication is a second-class citizen within browsers.
Future research should examine how browsers can better support password-based authentication---for example, making password-based authentication interfaces first-class HTML elements that the browser implements to ensure that passwords are handled correctly.
This could include providing a common, recognizable interface for password-based authentication, allowing for the use of alternative protocols (e.g., strong password protocols~\cite{bellovin1992encrypted,wu1998secure}), and preventing malicious websites from creating look-alike phishing interfaces~\cite{ruoti2017end}.

Research should also explore how browsers can provide additional features to password manager extensions.
Examples include,
(1) allowing password managers to generate a nonce to autofill in place of the password that the browser will replace with the password when it is transmitted to the website if and only if the target domain matches the domain associated with the password in the password manager~\cite{stock2014protecting} (see Section~\ref{sec:mitigation});
(2) providing password managers access to the system keyring (e.g., macOS keyring, Windows Vault), giving them a more secure and standardized mechanism for storing account credentials;
(3) handling the user interaction component of autofill and ensuring that it is clickjack resilient;
(4) adding HTML attributes that describe a website's password policy, allowing password managers to generate passwords that will be accepted by the website~\cite{stajano2014password}.

\textbf{Research-Derived Character Sets.}
Password managers generate passwords using different character sets, differing dramatically in which symbols they allow and which characters they remove as unusable (e.g., difficult to remember, hard to distinguish).
We advocate for a data-driven effort to establish standardized character sets.

User studies should be conducted to identify the characters that are difficult for users to read and input, with attention paid to alternative input modalities (e.g., entering passwords using a TV remote or accessible keyboard).
Measurements of existing password policies could also be used to identify which characters are commonly rejected by website password policies.
It may be that there is no one ideal character set, but rather different character sets for different types of passwords (e.g., passwords with restrictive policies, passwords entered with non-keyboard modalities).
In this case, statistical modeling could be used to identify the ideal lengths for passwords in various modalities.

\textbf{HTML-Supported Password Generation.}
Stajano et al.~\cite{stajano2014password} recommended adding HTML attributes to help password managers identify the policy to use when generating passwords.
We believe that this approach should receive more attention.
In particular, it would be helpful to see developer studies studying the feasibility adding this feature to existing websites and user studies to ensure that this feature is understandable and helpful to users.
It would also be worth examining whether such annotations could be automatically inferred and added by semantically evaluating the code that checks passwords.

\textbf{Mobile Password Managers.}
Our work examined the security of password managers in a desktop environment.
Given the prevalence of mobile devices, a similar analysis of the security of mobile password managers is necessary.

\section{Conclusion}
Password managers are currently being recommended by the media~\cite{cnet,pcmag}; as such, it is disappointing that users need to be cautious when selecting a password manager and must also spend time to ensure that they understand how to correctly configure it.
As experience has shown, pushing these responsibilities onto users rarely has the expected outcome~\cite{herley2009so}.
Therefore, we believe it is important that researchers continue to evaluate the progress of password managers---both in terms of security and usability---and that work is done to continue to improve the security and usability of password managers~\cite{ruoti2017end}.

\section*{Disclosure}
We have made these results available to the maintainers of each password manager studied. RoboForm has already adopted several of our recommendations.

\section*{Research Artifacts}
The generated data, scripts used to analyze that data, and all analysis artifacts are available for download at \url{https://userlab.utk.edu/papers/oesch2020that}.

\section*{Acknowledgments}
The authors would like the thank their shepherd Ben Stock and the anonymous reviewers for their helpful feedback. 

\balance
\bibliographystyle{plain}
\bibliography{main}

\cleardoublepage
\onecolumn
\appendix

\section{Additional Password Generation Data}
\label{appx:generation}

\begin{appxitem}
	\setuptable
	
	\rowcolors{3}{}{gray!10}
	\begin{tabular}{l|ll|ll|ll|ll|ll|}
		& \multicolumn{2}{c|}{all}
		& \multicolumn{2}{c|}{l}
		& \multicolumn{2}{c|}{ld}
		& \multicolumn{2}{c|}{ls}
		& \multicolumn{2}{c|}{sd} \\
		
		System
		& \multicolumn{1}{c}{$p$} & \multicolumn{1}{c|}{$\chi^2$}
		& \multicolumn{1}{c}{$p$} & \multicolumn{1}{c|}{$\chi^2$}
		& \multicolumn{1}{c}{$p$} & \multicolumn{1}{c|}{$\chi^2$} 
		& \multicolumn{1}{c}{$p$} & \multicolumn{1}{c|}{$\chi^2$}
		& \multicolumn{1}{c}{$p$} & \multicolumn{1}{c|}{$\chi^2$} \\ \hline
		
		KeePassX		& 1.00 & 84.62 & 1.00 & 42.15 & 1.00 & 65.49 & 1.00 & 77.38 & 1.00 & 38.81	\\
		KeePassXC		& 1.00 & 85.16 & 1.00 & 67.35 & 1.00 & 61.41 & 1.00 & 76.88 & 1.00 & 35.27	\\ \hline
		
		1Password X & \sig 0.00 & \sig 294756		& 1.00 & 41.80 & \sig 0.00 & \sig 132469  & \sig 0.00 & \sig 17747		&				&	\\
		Bitwarden		& \sig 0.00 & \sig 724697		& 1.00 & 53.40 & \sig 0.00 & \sig 361209  & \sig 0.00 & \sig 362807		& 1.00	& 12.54		\\
		Dashlane		& \sig 0.00 & \sig 729301		& \sig 0.00 & \sig 1203  & \sig 0.00 & \sig 334844  & \sig 0.00 & \sig 47489		& \sig 0.00	& \sig 348990	\\
		LastPass		& \sig 0.00 & \sig 640316		& 1.00 & 72.20 & \sig 0.00 & \sig 96928 	& \sig 0.00 & \sig 390413		& \sig 0.00	& \sig 156327	\\
		RoboForm		& \sig 0.00 & \sig 1108211	& \sig 0.00 & \sig 10792 & \sig 0.00 & \sig 470973 	& \sig 0.00 & \sig 605343		& \sig 0.00	& \sig 41584		\\ \hline
		
		Chrome			& 1.00	& 54.95 & 1.00 & 38.50   & 1.00 & 47.51      & 1.00 & 40.28 & 1.00 & 16.16 \\ \hline
		
		SPG				& \sig 0.00 & \sig 445079 & 1.00 & 45.67 & \sig 0.00 & \sig 245539 & \sig 0.0 & \sig 10804 & \sig 0.0 & \sig 190506 \\
		\texttt{/dev/rand}	& 1.00 & 77.65 & 1.00 & 59.37 & 1.00 & 62.17 & 1.00 & 89.01 & 1.00 & 37.73 \\ \hline
		
	\end{tabular}
	
	\captionof{figure}{Length 8 $\chi^{2}$ Scores for Character Frequency}
	\label{tab:chisquaredvalues8}
\end{appxitem}

\begin{appxitem}
	\setuptable
	
	\rowcolors{3}{white}{gray!10}
	\begin{tabular}{l|ll|ll|ll|ll|ll|}
		& \multicolumn{2}{c|}{all}
		& \multicolumn{2}{c|}{l}
		& \multicolumn{2}{c|}{ld}
		& \multicolumn{2}{c|}{ls}
		& \multicolumn{2}{c|}{sd} \\
		
		System
		& \multicolumn{1}{c}{$p$} & \multicolumn{1}{c|}{$\chi^2$}
		& \multicolumn{1}{c}{$p$} & \multicolumn{1}{c|}{$\chi^2$}
		& \multicolumn{1}{c}{$p$} & \multicolumn{1}{c|}{$\chi^2$} 
		& \multicolumn{1}{c}{$p$} & \multicolumn{1}{c|}{$\chi^2$}
		& \multicolumn{1}{c}{$p$} & \multicolumn{1}{c|}{$\chi^2$} \\ \hline
	
		KeePassX		& .65 & 87.09 & .74 & 44.12 & .03 & 84.43 & .45 & 83.96 & .11 & 52.57\\
		KeePassXC		& .052 & 116.17 & .44 & 51.78 & .56 & 58.64 & .65 & 77.46 & .54 & 39.42\\ \hline
		
		1Password X     & \sig0.00 &\sig 95480  & .54 & 45.44 & \sig0.00 & \sig33175  & \sig0 & \sig1600		&				&	\\
		Bitwarden		& \sig0.00 & \sig481688  & .49 & 48.60 & \sig0.00 & \sig239474  & \sig0.00 & \sig241181  & .21 & 19.20\\
		Dashlane		& \sig0.00 & \sig487295  & \sig0.00 & \sig765  & \sig0.00 & \sig224131 & \sig0.00 & \sig32113  & \sig0.00 & \sig233758\\
		LastPass		& \sig0.00 & \sig428916 & .73 & 44.30 & \sig0.00 & \sig64703 & \sig0.00 & \sig258080 & \sig0.00 & \sig104851\\
		RoboForm		& \sig0.00 & \sig738458 &\sig 0.00 & \sig7277 & \sig0.00 & \sig312865 & \sig0.00 & \sig403972 & \sig0.00 & \sig27661\\ \hline
		
		Chrome			& .70 & 53.71     & .51 & 46.11   & .15 & 65.53      & .99 & 31.27 & \sig0.00 & \sig34.3\\ \hline
		
		Web generator				& \sig0.00 & \sig297694 & \sig.047 & \sig69.07 & \sig0.00 & \sig163675 & \sig0.00 & \sig7289 & \sig0.00 & \sig125531 \\
		\texttt{/dev/rand}	        & .33 & 99.23 & .27 & 56.73 & .75 & 53.10 & .31 & 89.93 & .55 & 40.11\\ \hline

	\end{tabular}

	\captionof{table}{Length 12 $\chi^{2}$ Scores for Character Frequency}
	\label{tab:chisquaredvalues12}
\end{appxitem}

\begin{appxitem}
	\setuptable
	
	\rowcolors{3}{white}{gray!10}	
	\begin{tabular}{l|ll|ll|ll|ll|ll|}
		& \multicolumn{2}{c|}{all}
		& \multicolumn{2}{c|}{l}
		& \multicolumn{2}{c|}{ld}
		& \multicolumn{2}{c|}{ls}
		& \multicolumn{2}{c|}{sd} \\
		
		System
		& \multicolumn{1}{c}{$p$} & \multicolumn{1}{c|}{$\chi^2$}
		& \multicolumn{1}{c}{$p$} & \multicolumn{1}{c|}{$\chi^2$}
		& \multicolumn{1}{c}{$p$} & \multicolumn{1}{c|}{$\chi^2$} 
		& \multicolumn{1}{c}{$p$} & \multicolumn{1}{c|}{$\chi^2$}
		& \multicolumn{1}{c}{$p$} & \multicolumn{1}{c|}{$\chi^2$} \\ \hline
	
		KeePassX		& .62 & 88.10 & .91 & 38.06 & .14 & 73.07 & .11 & 98.34 & .30 & 45.11\\
		KeePassXC		& .49 & 92.57 & .79 & 42.76 & .97 & 41.66 & .71 & 75.38 & .92 & 28.97\\ \hline
		
		1Password X     & \sig0.00 & \sig12789  & .82 & 38.21 & \sig0.00 & \sig2367  & \sig.03 & \sig90.32		&				&	\\
		Bitwarden		& \sig0.00 & \sig289893  & .72 & 42.84 & \sig0.00 & \sig143389  & \sig0.00 & \sig143720  & .21 & 19.10\\
		Dashlane		& \sig0.00 & \sig956201  & \sig0.00 & \sig443060  & \sig0.00 & \sig401737  & \sig0.00 & \sig822537  & .17 & 42.48\\
		LastPass		& \sig0.00 & \sig256787 & .50 & 50.32 & \sig0.00 & \sig38336 & \sig0.00 & \sig156177 & \sig0.00 & \sig63559\\
		RoboForm		& \sig0.00 & \sig442762 & \sig0.00 & \sig4524 & \sig0.00 & \sig188292 & \sig0.00 & \sig241760 & \sig0.00 & \sig16928\\ \hline
		
		Chrome			& .91 & 46.01     & .36 & 49.88   & .25 & 61.8      & .50 & 51.2 & \sig.056 & \sig20.60\\\hline
		
		Web generator				& \sig0.00 & \sig178091  & .69 & 45.53 & \sig0.00 & \sig98651 & \sig0.00 & \sig4617 & \sig0.00 & \sig75043 \\
		\texttt{/dev/rand}	        & .63 & 88.73 & .22 & 58.42 & .29 & 66.77 & .24 & 92.88 & .49 & 41.62\\ \hline

	\end{tabular}
	
	\captionof{table}{Length 20 $\chi^{2}$ Scores for Character Frequency}
	\label{tab:chisquaredvalues20}
\end{appxitem}

\begin{appxitem}
	\setuptable
	
	\rowcolors{2}{gray!10}{}
	\begin{tabular}{ll|l|}
		Length & Composition & \multicolumn{1}{c|}{Characters Sorted by Frequency} \\ \hline
		
		8		& all	& \verb|%!^@$#4627583NPHFDJUXACTSGMERBKLQVYWqmgasneokfvptbuhyixdwrcjzZ9| \\
		12	& all	& \verb|$%#^!@2637548BHGFSQECXWYJRDNMUALVPTKdtboenhskjvqaicgwpmxfyur9zZ| \\
		20	& all	& \verb|%#@!^$8462735XHVPJWCUFKLYNDESAMTQRiBGgdveaspnkytqjfbmxwrcuoh9Zz| \\ \hline
		8		& l		& \verb|GHDYEQKPJFURCTASNLVMXBWpyikuvmtofxecasdwjngbhqrZ| \\
		12	& l		& \verb|VMDFQAGNRLUEXKCJSBPWTYHcmfiqyawnektsdvrgjhopxbuZ| \\
		20	& l		& \verb|REFQWJUTBKDGCMAHSVPXYLNfkvyjsnhwoepabqixgdturcmZ| \\ \hline
		8		& ld	& \verb|5782346RUALJDQFHSPKEVGTMYBXCNWhynabrqwpkfumxjvctodsigeZ9| \\
		12	& ld	& \verb|6853247JUWYSBLTQFGCRMPVKANXHEDgcidbjtwpesafxqvhmrkounyZ9| \\
		20	& ld	& \verb|6532874MTJFSVCYDNHPLGWEXQABUnRkeKswpjughytdqbircafovxm9Z| \\ \hline
		8		& ls	& \verb|%@^$#!SLFWVAURKNTEXDQJYBMHPGCavhtndwcjkyufxieqobrgpmszZ| \\
		12	& ls	& \verb|$^%#@!FHJVESBGMUYXDLTPCAQNWRKrwogjhicexmsyftvkqdabupnZz| \\
		20	& ls	& \verb|%@$^#!PFAXTKBQCSHDGVJEMWRYtNgUfLabyshrkpwmdouvqxjineczZ| \\ \hline
		8		& sd	& \verb|#$@%!^65324879| \\
		12	& sd	& \verb|$@!#^%57263489| \\
		20	& sd	& \verb|@^!$#%63582749| \\ \hline
		
	\end{tabular}

	\captionof{table}{Character Frequencies of Generated Passwords from RoboForm}
	\label{tab:robocharfreq}
\end{appxitem}

\begin{appxitem}
	\begin{minipage}{.425\textwidth}
	  \begin{lstlisting}[language=json,firstnumber=1]
{
 "args": {
   "pwd_file": ["$TRAINING_FILE"],
   "pwd_format": ["list"],
   "log_file": "$LOG_FILE",
   "arch_file": "$ARCH_FILE",
   "weight_file": "$WEIGHT_FILE"
 },
 
 "config": {
   "intermediate_fname": "$INTERMEDIATE_FILE",
   "min_len": $PASSWORD_LENGTH,
   "max_len": $PASSWORD_LENGTH,
   
   "training_chunk": 1024,
   "layers": 2,
   "hidden_size": 1000,
   "dense_layers": 1,
   "dense_hidden_size": 512,
   "generations": 5
 }
}
		\end{lstlisting}
	\end{minipage}%
	\hspace{.05\textwidth}%
	\begin{minipage}{.525\textwidth}
		\begin{lstlisting}[language=json,firstnumber=1]
{
 "args": {
   "enumerate_ofile": "$GUESSES_FILE",
   "log_file": "$LOG_FILE",
   "arch_file": "$ARCH_FILE",
   "weight_file": "$WEIGHT_FILE"
 },

 "config": {
   "guess_serialization_method": "delamico_random_walk",
   "password_test_fname": "$TESTING_FILE",
   "parallel_guessing": true,
   
   "intermediate_fname": "$INTERMEDIATE_FILE",
   "min_len": $PASSWORD_LENGTH,
   "max_len": $PASSWORD_LENGTH,
   
   "training_chunk": 1024,
   "layers": 2,
   "hidden_size": 1000,
   "dense_layers": 1,
   "dense_hidden_size": 512,
   "generations": 5
 }
}
		\end{lstlisting}
	\end{minipage}
	\vspace{-.7em}
	\captionof{lstlisting}{Neural Network Configuration---Training (Left) and Testing (Right)}
	\label{lst:nnconfig}
\end{appxitem}

\end{document}